\documentclass[11pt]{article}
\usepackage{jheppub}
\usepackage{bm,bbm}
\usepackage{booktabs,amsmath}
\usepackage{accents}
\usepackage{multirow}
\usepackage{graphics}
\usepackage{braket}
\usepackage{mathtools}
\usepackage{comment}
\usepackage{autobreak}
\usepackage{caption,subcaption}
\usepackage{enumerate}
\usepackage{longtable}
\usepackage{enumitem}
\usepackage[dvipsnames]{xcolor}
\usepackage[scr=stixtwoplain]{mathalpha}
\usepackage{tabularray}
\UseTblrLibrary{booktabs}

\usepackage{nicematrix}

\usepackage{tikz}

\usetikzlibrary{patterns}
\usetikzlibrary{arrows,shapes,positioning}
\usetikzlibrary{decorations.markings}
\usetikzlibrary{decorations.pathmorphing}
\tikzset{snake it/.style={decorate, decoration=snake}}

\tikzset{->-/.style 2 args={
    postaction={decorate},
    decoration={markings, mark=at position #1 with {\arrow[thick, #2]{>}}}
    },
    ->-/.default={0.7}{}
}

\tikzset{-<-/.style 2 args={
    postaction={decorate},
    decoration={markings, mark=at position #1 with {\arrow[thick, #2]{<}}}
    },
    -<-/.default={0.7}{}
}

\hypersetup{
    colorlinks=true,
    linkcolor=blue,
    filecolor=magenta,      
    urlcolor=cyan,
    pdfpagemode=FullScreen,
}

\usepackage{arydshln}
\usepackage{mathtools}
\edef\restoreparindent{\parindent=\the\parindent\relax}
\usepackage{parskip}
\restoreparindent

\usepackage{amsthm}
\newtheoremstyle{break}
  {\topsep}{\topsep}%
  {\upshape}{}%
  {\bfseries}{}%
  {\newline}{}%
\theoremstyle{break}

\def\Tr{{\rm Tr}}
\def\d{{\rm d}}
\def\i{{\rm i}}

\def\CB{{\cal B}}

\def\CH{{\cal H}}

\def\CL{{\cal L}}

\def\CN{{\cal N}}
\def\CO{{\cal O}}

\def\BF{\mathbb{F}}
\def\BN{\mathbb{N}}

\def\BR{\mathbb{R}}

\def\BZ{\mathbb{Z}}

\def\SC{\mathsf{C}}
\def\SG{\mathsf{G}}
\def\SH{\mathsf{H}}

\def\SCZ{\mathscr{Z}}
\def\d{\mathrm{d}}
\def\SO{\mathrm{SO}}

\def\U{\mathrm{U}}

\title{
Gauging $\BZ_N$ symmetries of Narain CFTs
}

\author[a]{Keiichi Ando,} 
\author[b]{Kohki Kawabata,}
\author[a]{and Tatsuma Nishioka}

\affiliation[a]{Department of Physics, The University of Osaka,\\
Machikaneyama-Cho 1-1, Toyonaka 560-0043, Japan}

\affiliation[b]{Department of Physics, Faculty of Science,
The University of Tokyo,\\
Bunkyo-Ku, Tokyo 113-0033, Japan}

\preprint{OU-HET-1269}

\abstract{
We investigate the gauging of a $\BZ_N$ symmetry in lattice conformal field theories (CFTs), also known as Narain CFTs.
For prime $N$, we derive a spin selection rule for operators in a $\BZ_N$ charge-twisted sector of a general bosonic CFT.
Using this result, we formulate the gauging procedures in lattice CFTs as modifications of the momentum lattices by a lattice vector that specifies a non-anomalous $\BZ_N$ symmetry.
Applying this formulation to code CFTs, i.e., Narain CFTs constructed from error-correcting codes, we express the torus partition functions of the orbifolded and parafermionized theories in terms of the weight enumerator polynomials of the underlying codes.
As an application, we identify a class of codes that yield self-dual bosonic CFTs under the orbifolding by a $\BZ_N$ symmetry.
}

\begin{document} 
\maketitle
\flushbottom

\newpage

\section{Introduction}
Conformal field theories (CFTs) play a pivotal role in describing the universal features of critical phenomena in both high energy theory and condensed matter systems.
In two dimensions, conformal symmetry is enhanced to the infinite-dimensional Virasoro algebra and restricts the forms of the correlation functions \cite{1984NuPhB241333B,ginsparg1988applied-436,DiFrancesco:1997nk}.
This rich symmetry structure allows for the exact computation of critical exponents and the classification of critical points in statistical mechanics.
Moreover, the modular invariance of the torus partition function leads to a set of constraints on the spectrum of the operators, allowing for the classification of partition functions in various bosonic theories \cite{Cardy1986ie,Cappelli:1986hf,Cappelli:1987xt,Kato:1987td,DiFrancesco:1997nk}.

CFTs with fermions are of significant importance in the study of superstring theory \cite{Polchinski:1998rr}. 
They also provide a framework for understanding symmetry-protected topological (SPT) phases and other exotic phases of matter in condensed matter physics (see e.g., \cite{fidkowski2011topological-a6b,kapustin2015fermionic-27b}).
Fermionic CFTs have the fermion parity $\BZ_2$ symmetry and depend on the spin structure of Riemann surfaces which specify the periodicity of a fermionic operator along a non-contractible cycle.
Since the spin structure can change under modular transformations, the partition functions of fermionic theories are not modular invariant but rather transform covariantly.
It is, however, possible to gauge the fermion parity $\BZ_2$ symmetry and construct modular invariant theories from fermionic ones.
In string theory, this procedure is known as the GSO projection \cite{gliozzi1977supersymmetry-ebe}, while in a broader context, it is referred to as bosonization, as the resulting theories are bosonic.
Conversely, when bosonic theories have a non-anomalous $\BZ_2$ symmetry they can be mapped to their fermionic counterparts by fermionization.
In a modern viewpoint, fermionization is described by stacking a bosonic CFT with the Kitaev Majorana chain in the $\BZ_2$ non-trivial topological phase \cite{Kitaev:2000nmw} followed by gauging the diagonal $\BZ_2$ symmetry \cite{gaiotto2016spin-eb1,kapustin2017fermionic-f9b,Tachikawa-TASI,Karch:2019lnn,Ji:2019ugf,thorngren2020anomalies-227}.
Fermionization has been applied to classify some classes of fermionic CFTs such as fermionic minimal models~\cite{Hsieh:2020uwb,Kulp:2020iet} and chiral fermionic CFTs with relatively small central charges~\cite{BoyleSmith:2023xkd,Hohn:2023auw,Rayhaun:2023pgc}.

Gauging operations are not limited to $\BZ_2$ symmetries but can be extended to $\BZ_N$ symmetries, provided these symmetries are free of anomalies.
A typical example of $\BZ_N$-gaugings is orbifolding, which projects the theory onto its $\BZ_N$-invariant sector while simultaneously introducing twisted sectors associated with non-trivial symmetry elements and modifying the spectrum and operator content.
A more intricate form of $\BZ_N$-gauging is parafermionization, which extends the fermionization procedure.
This operation maps a bosonic CFT to a parafermionic CFT, which hosts operators with fractional spins and has a $\BZ_N$ symmetry analogous to the fermion parity in fermionic theories \cite{Fateev:1985mm,zamolodchikovnoyearnonlocal-3de,Gepner:1986hr}.
The inverse procedure, bosonization, can also be achieved by gauging the $\BZ_N$ symmetry of parafermionic CFTs.
The structure of $\BZ_N$-gaugings in two-dimensional CFTs have been explored systematically in recent studies \cite{Chang:2018iay,Lin:2021udi,Thorngren:2021yso,chang2023topological-18b,chen2023para-fusion-0eb} by using topological defect lines for describing symmetry actions and gauging procedures.
The bosonization/parafermionization map on a torus has been formulated explicitly and investigated, elucidating the modular properties of the partition functions and the duality relations between the theories \cite{Yao:2020dqx,Thorngren:2021yso,Duan:2023ykn}.

In this paper, we study $\BZ_N$-gaugings in CFTs of Narain type \cite{Narain:1985jj,Narain:1986am}, which possess $\BZ_N$ symmetries as subgroups of the shift and winding $\U(1)$ symmetries, and systematically construct the orbifolded and parafermionized theories on a torus.
This work is motivated by the extension of our previous approach for $\BZ_2$-gaugings \cite{Kawabata:2023iss,Kawabata:2023usr,Ando:2024gcf} to $\BZ_N$-gaugings within the context of code CFTs, a special class of Narain-type lattice CFTs constructed from classical and quantum error correcting codes \cite{frenkel1984natural,frenkel1989vertex,Dolan:1994st,Gaiotto:2018ypj,Kawabata:2023nlt,Kawabata:2023rlt,Kawabata:2024gek,Okada:2024imk,Dymarsky:2020bps,Dymarsky:2020qom,Dymarsky:2020pzc,Dymarsky:2021xfc,Henriksson:2021qkt,Buican:2021uyp,Yahagi:2022idq,Furuta:2022ykh,Henriksson:2022dnu,Angelinos:2022umf,Henriksson:2022dml,Dymarsky:2022kwb,Kawabata:2022jxt,Furuta:2023xwl,Alam:2023qac,Kawabata:2023usr,Kawabata:2023iss,Aharony:2023zit,Barbar:2023ncl,Singh:2023mom,Ando:2024gcf,Singh:2024qjm,Mizoguchi:2024ahp,Kawabata:2025hfd}.
The $\BZ_2$-gaugings of code CFTs have proven to be instrumental in searching for bosonic CFTs invariant under the orbifolding as well as fermionic CFTs with supersymmetry from code perspective \cite{Kawabata:2023iss,Kawabata:2023usr,Ando:2024gcf}.
Building on these successes, we aim to establish a general methodology for systematically constructing new classes of bosonic and parafermionic CFTs from lattice CFTs by $\BZ_N$-gauging operations.

To achieve this purpose, we first revisit gauging in general bosonic CFTs with a non-anomalous $\BZ_N$ symmetry on a torus. 
We use topological defect lines to describe the modular transformation law of the torus partition functions in the twisted sectors.
We then derive a spin selection rule for the operators with a fixed $\BZ_N$ charge in the twisted sectors. 
This result can be seen as a generalization of a similar rule for a non-anomalous $\BZ_2$ symmetry in \cite{Lin:2019kpn}.
Since the spin selection rule simplifies for prime $N$, we focus on this case throughout this paper.
We determine how the Hilbert spaces of charge-twisted sectors are shuffled under the $\BZ_N$-gauging operations by examining the transformation laws of the torus partition functions under orbifolding and parafermionization.
In addition, we obtain the spin selection rules in the orbifolded and parafermionized theories.

Next, we reformulate the $\BZ_N$-gauging operations in lattice CFTs as modifications of the associated Narain momentum lattices. 
In this setting, the $\BZ_N$ symmetry is specified by a lattice vector encoding a subgroup of the shift and winding $\U(1)$ symmetries.
We derive the condition for this symmetry to be non-anomalous in terms of the lattice vector and express the torus partition functions of the orbifolded and parafermionized theories in terms of the modified momentum lattices.
Furthermore, for code CFTs, we show that the torus partition functions of the $\BZ_N$-gauged theories can be written by the weight enumerator polynomials of the underlying codes, with their variables replaced by certain combinations of theta functions that depend on the gauging data.
This formulation enables us to construct explicit examples of parafermionic CFTs from codes, which accommodate operators with fractional spins consistent with our spin selection rules.

In addition to illustrating the $\BZ_N$-gauging procedures in code CFTs by numerous examples, we use our results to construct self-dual bosonic CFTs under the orbifolding from Calderbank-Shor-Steane (CSS) codes that are characterized by a pair of classical codes $\SC_\textsf{X}$ and $\SC_\textsf{Z}$ satisfying a certain condition \cite{Calderbank:1995dw,steane1996multiple}.
We identify a class of classical codes $\SC$ such that code CFTs built from CSS codes with $\SC_\textsf{X} = \SC^\perp$ and $\SC_\textsf{Z} = \SC$ become self-dual under the orbifolding by a $\BZ_N$ symmetry.
Self-dual theories have played a key role in the studies of non-invertible symmetries and duality defects \cite{Choi:2021kmx,Shao:2023gho}, and our results may have potential applications in this direction.

The remainder of the paper is structured as follows.
In section \ref{ss:ZN_bosonic}, we revisit $\BZ_N$-gaugings in general bosonic CFTs on a torus and derive the spin selection rule for operators in a charge-twisted sector.
We then read off the mapping rules of the charge-twisted sectors under the orbifolding and parafermionization from the transformation laws of the torus partition functions under the gaugings.
Section \ref{ss:ZN-gauging-lattice-CFTs} presents our general formulation of $\BZ_N$-gaugings in lattice CFTs, where we describe gauging as modifications of the momentum lattices by a lattice vector corresponding to a non-anomalous $\BZ_N$ symmetry.
In section \ref{ss:codeCFT}, we apply our formulation to code CFTs and construct the orbifolded and parafermionized theories.
We derive the explicit forms of the torus partition functions given by the weight enumerator polynomial of the associated code.
In section \ref{ss:Examples}, our constructions are illustrated with numerous examples.
Section \ref{ss:self-dual} explores a class of codes that give rise to self-dual code CFTs under the orbifolding by a $\BZ_N$ symmetry.
Finally, section \ref{ss:discussion} summarizes our findings and outlines possible directions for future work.

\section{$\BZ_N$-gauging of bosonic CFTs}\label{ss:ZN_bosonic}
In this section, we consider $\BZ_N$-gaugings of a two-dimensional bosonic CFT on a torus.
Our discussion closely follows the previous works \cite{Chang:2018iay,Lin:2021udi} which employ topological defect lines for describing a $\BZ_N$ symmetry and identifying the modular transformation laws of the torus partition function.
In section \ref{ss:twisted_sector}, we introduce the Hilbert spaces in the twisted sectors by inserting the topological defect lines along the temporal direction.
Focusing on the case with prime $N$, we derive the spin selection rule for operators in the twisted sectors with a fixed $\BZ_N$-charge from the transformation laws of the partition function under the modular $T$ transformation.
In section \ref{ss:orf_paraferm_ZN_gauging_general}, we rewrite the orbifolding and parafermionization procedures given in terms of the torus partition functions \cite{Yao:2020dqx,Thorngren:2021yso,Duan:2023ykn} as the mapping rules of the Hilbert spaces in the charge-twisted sectors.
In passing, we determine the spin selection rules for operators in the orbifolded and parafermionized theories.

\subsection{Twisted and untwisted sectors by $\BZ_N$ symmetry}\label{ss:twisted_sector}

\begin{figure}
    \centering
    \begin{subfigure}[b]{0.4\textwidth}
        \begin{tikzpicture}[scale=2.0,
        thick, >=stealth]
            \draw (0,0)--(2,0)--(2,2)--(0,2)--(0,0);
            \draw[RoyalBlue, ->-={.5}{scale=1.75}, very thick] (0,1) -- (2,1);
            \node[left=0.25cm] at (0,1) {\Large $g$};
            \node at (1, -0.25) {\large $\CH_0$};
            \node at (1, -0.75) {\large (a)};
        \end{tikzpicture}
    \end{subfigure}
    \qquad\qquad
    \begin{subfigure}[b]{0.4\textwidth}
        \begin{tikzpicture}[scale=2.0,
        thick, >=stealth]
            \draw (0,0)--(2,0)--(2,2)--(0,2)--(0,0);
            \node at (1, -0.25) {\large $\CH_b$};
            \draw[RoyalBlue, ->-={.5}{scale=1.75}, very thick] (0.75,0) --+ (0,2);
            \draw[RoyalBlue, ->-={.5}{scale=1.75}, very thick] (1.25,0) --+ (0,2);
            \node[RoyalBlue] at (1, 0.9) {\large $\cdots$};
            \node[above, black] at (1,2) {\Large $g^b$};
            \node at (1, -0.75) {\large (b)};
        \end{tikzpicture}
    \end{subfigure}
    \caption{Topological defect lines inserted along the spacial circle (a) and the temporal circle (b), respectively. The left panel (a) shows the action of $g$ on the untwisted sector, while the right panel (b) shows the definition of the $b$-th twisted sector $\CH_b$.}
    \label{fig:TDL}
\end{figure}
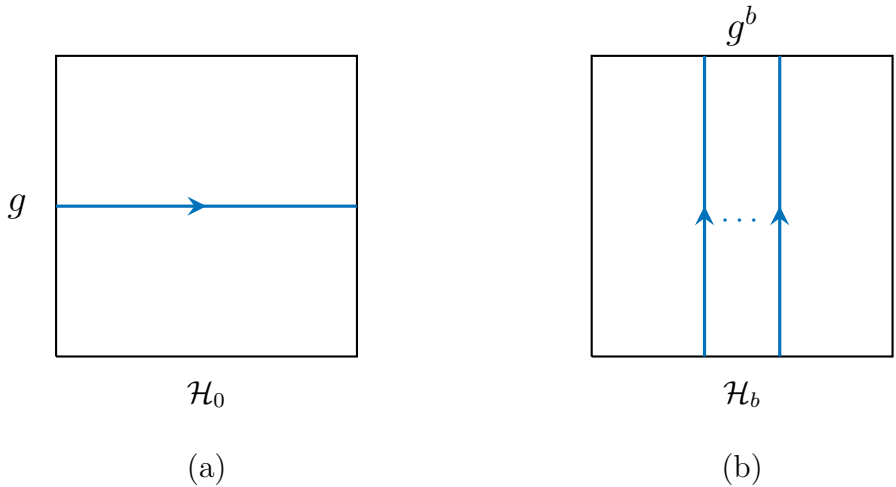

We consider a two-dimensional bosonic CFT $\CB$ with a non-anomalous global $\BZ_N$ symmetry generated by $g$.
Let $\CH_0$ be the Hilbert space of local operators in the theory $\CB$.
The Hilbert space $\CH_0$ can be graded under the $\BZ_N$ symmetry as
\begin{align}
    \CH_0
        =
            \bigoplus_{a\in \BZ_N} \CH_0^a \ ,
\end{align}
where the subsector $\CH_0^a$ carries the $\BZ_N$ charge $a\in \BZ_N = \{ 0, 1, \cdots, N-1\}$:
\begin{align}
    g\, \varphi = \omega_N^a\,\varphi \ , \qquad \varphi \in \CH_0^{a}  \ ,
\end{align}
where $\omega_N$ is the $N$-th root of unity, $\omega_N = e^{\frac{2\pi\i}{N}}$.
The action of the generator $g$ can be represented by the insertion of the topological defect line along the spacial direction in the path integral picture as shown in figure \ref{fig:TDL}\,(a).

We can extend the Hilbert space $\CH_0$ by introducing the twisted sector $\CH_b~(b=1, 2, \cdots, N-1)$, which is the Hilbert space quantized under the twisted boundary condition $\varphi(\sigma+2\pi) = \omega_N^b\,\varphi(\sigma)$ for a bosonic field $\varphi$ with unit $\BZ_N$ charge when the theory is located on a circle parametrized by $\sigma\in [0, 2\pi)$.
The twisted boundary condition in the sector $\CH_b$ can also be described by inserting the topological defect lines $b$ times along the temporal direction as shown in figure \ref{fig:TDL}\,(b).

The $b$-th twisted sector $\CH_b$ can be graded in a similar manner to $\CH_0$ under the $\BZ_N$ symmetry as
\begin{align}\label{decomposition_hilbert_spaces}
    \CH_b
        =
            \bigoplus_{a\in \BZ_N} \CH_b^a \ ,
\end{align}
where the subsector $\CH_b^a$ carries the $\BZ_N$ charge $a\in \BZ_N$ in the absence of an anomaly.
For a state with conformal weight $(h, \bar h)$ in $\CH_b$, the spin $s:= h - \bar h$ satisfies
\begin{align}\label{Non-anomalous-condition-ZN}
    s = h - \bar h \in \frac{\BZ}{N} \ .
\end{align}
It has been shown in \cite{Chang:2018iay} that the above relation no longer holds when the $\BZ_N$ symmetry is anomalous (see also \cite{Hung:2013cda,Lin:2021udi}).
One can diagnose the anomaly of the $\BZ_N$ symmetry by checking whether \eqref{Non-anomalous-condition-ZN} holds.
In what follows, we will call \eqref{Non-anomalous-condition-ZN} the non-anomalous condition.

Now, consider the bosonic CFT $\CB$ on a torus with modulus $\tau$.
We couple $\CB$ with the $\BZ_N$ background gauge fields by introducing the holonomies $\alpha, \beta\in \BZ_N$ along the temporal and spacial circles, respectively.
Then, the Hilbert space becomes the $\beta$-twisted sector $\CH_\beta$ due to the holonomy $\beta$ and the bosonic partition function $\SCZ_\CB[\alpha, \beta]$ can be written as
\begin{align}\label{bosonic_PF}
    \SCZ_\CB[\alpha, \beta]
        &=
            \Tr_{\CH_{\beta}}\left[ g^\alpha\,q^{L_0 - \frac{c}{24}}\,\bar q^{\bar L_0 - \frac{\bar c}{24}}\right] \ ,
\end{align}
where $q := e^{2\pi\i\tau}$.
Here, $L_0$ is the zero mode of the Virasoro algebra with central charge $c$ for the left-moving part.
The symbols with the bar are for the right-moving part.

In general, the modular $T$ transformation changes the configuration of topological defect line operators as follows:
\begin{align}
\label{T_transformation}
    T\quad
    \begin{tikzpicture}[scale=1.5,baseline={(0,1.25)}, >=stealth, thick]
        \draw (0,0)--(2,0)--(2,2)--(0,2)--(0,0);
        \draw[very thick, orange, ->-={.7}{scale=1.5}] (1.4,0) -- (1.4,2);
        \node[below, orange] at (1.4,0) {$g'$};
        \draw[very thick, RoyalBlue, ->-={.5}{scale=1.5}] (0,1) -- (2,1);
        \node[left, RoyalBlue] at (0,1) {$g$};
    \end{tikzpicture}
    \qquad
    =
    \quad
     \begin{tikzpicture}[scale=1.5,baseline={(0,1.25)}, >=stealth, thick]
        \draw (0,0)--(2,0)--(2,2)--(0,2)--(0,0);
        \draw[very thick, orange, ->-={.7}{scale=1.5}] (1.4,0) -- (1.4,2);
        \node[below, orange] at (1.4,0) {$g'$};
        \draw[very thick, RoyalBlue, ->-={.5}{scale=1.5}] (0,1) -- (2,1);
        \node[left,RoyalBlue] at (0,0.9) {$g\,g'$};
        \draw[very thick, RoyalBlue, ->-={.5}{scale=1.5}] (0,0.8) -- (2,0.8);
    \end{tikzpicture}
\end{align}
On the left-hand side, $g, g'$ are topological defect lines along the spacial and temporal circles, respectively.
For simplicity, we denote \eqref{T_transformation} as 
\begin{align}
\label{T_rule}
    T\ (g,g')=(g\,g',g')\ .
\end{align}
By using this relation, one can obtain a non-trivial relation between the spin $s$ and the $\BZ_N$ charge $a$ of a state in the $b$-th twisted sector $\CH_b^a$ as follows.

First, we will determine an integer $n_b\in\BZ_N$ that satisfies the relation:
\begin{align}
\label{b_Twisted}
T^{n_b}\quad
    \begin{tikzpicture}[scale=1.5,baseline={(0,1.25)}, >=stealth, thick]
        \draw (0,0)--(2,0)--(2,2)--(0,2)--(0,0);
        \draw[very thick, orange, ->-={.7}{scale=1.5}] (0.6,0) -- (0.6,2);
        \draw[very thick, orange, ->-={.7}{scale=1.5}] (1.4,0) -- (1.4,2);
        \node[orange] at (1.0,1.0) {$\cdots$};
        \node[below, orange] at (1.0,0) {$\hat{\CL}^b$};
    \end{tikzpicture}
    \qquad
    &=
    \quad
     \begin{tikzpicture}[scale=1.5,baseline={(0,1.25)}, >=stealth, thick]
        \draw (0,0)--(2,0)--(2,2)--(0,2)--(0,0);
        \draw[very thick, orange, ->-={.7}{scale=1.5}] (0.6,0) -- (0.6,2);
        \draw[very thick, orange, ->-={.7}{scale=1.5}] (1.4,0) -- (1.4,2);
        \node[orange] at (1.0,1.0) {$\cdots$};
        \node[below, orange] at (1.0,0) {$\hat{\CL}^b$};
        \draw[very thick, RoyalBlue, ->-={.525}{scale=1.5}] (0,0.75) -- (2.0,0.75);
         \node[left, RoyalBlue] at (0,0.75) {$\hat{\CL}$};
    \end{tikzpicture}
\end{align}
Using the $T$ transformation law \eqref{T_rule}, the above relation \eqref{b_Twisted} can be written as
\begin{align}
        T^{n_{b}}\ (1,\hat{\CL}^b)
            =
                (\hat{\CL}^{n_b\,b},\hat{\CL}^b)
            =
                (\hat{\CL},\hat{\CL}^b) \ ,
\end{align}
which yields the following condition:
\begin{align}\label{condition_T}
    n_b\, b=1 \mod{N}\ .
\end{align}
The equation \eqref{condition_T} has a solution for any $b\in \BZ_N$ other than $b=0$ only when $N$ is a prime number.
In this case, using Fermat's little theorem we find the solution
\begin{align}
    n_b=b^{N-2} \mod{N}\ .
\end{align}

Next, we consider the action of $T^{n_b}$ and $\hat \CL$ on the Hilbert space $\CH^a_{b}$.
Since the $T$ transformation gives us the phase $e^{2\pi \i s}$ for a spin-$s$ state, the action of $T^{n_b}$ induces the phase $e^{2\pi \i n_b s}$ for the same state.
On the other hand, the topological defect line $\hat \CL$ acting on a state in $\CH_b^a$ with charge $a$ induces the phase $e^{2\pi \i \frac{a}{N}}$ when the $\BZ_N$ symmetry is non-anomalous.\footnote{In the presence of an anomaly, the charges of the twisted sectors become fractional.
For example, when $N=2$, the $\BZ_2$ charges take values in $a=\pm\tfrac{1}{2}$ rather than $a=0,1$. (See e.g., \cite{Tachikawa:2017gyf,Ji:2019ugf}.)}
Since the integer $n_b$ is chosen such that $T^{n_b}$ and $\hat \CL$ act in the same way, we find the non-trivial relation between the spin and charge of a state with charge $a$ in the $b$-th twisted Hilbert space $\CH_b^a$:
\begin{align}
    \label{graid_twisted}
        \begin{aligned}
        N\, n_b\,s&=a \mod N \ .
    \end{aligned}
\end{align}
Using the non-anomalous condition \eqref{Non-anomalous-condition-ZN} and \eqref{condition_T}, one can read off from \eqref{graid_twisted} the following spin selection rule:\footnote{For chiral theories, our spin selection rule leads to the selection rule for the conformal dimensions of operators obtained in chiral lattice VOAs \cite{mller2016cyclic-a51} (see also \cite{Burbano:2021loy}).}
\begin{align}\label{spin_selection_rule_bosonic}
    \CH_b^a\,:\qquad 
    s \in \BZ + \frac{a\,b}{N}  \ .
\end{align}
This spin selection rule is a generalization of the one for a non-anomalous $\BZ_2$ symmetry in \cite{Lin:2019kpn} to a non-anomalous $\BZ_N$ symmetry with arbitrary prime $N$.
The spin selection rules for $N=2$ and $N=3$ are shown in table \ref{tab:spin_selec}.

\begin{table}[]
    \centering
    \begin{subtable}[b]{0.2\textwidth}
    \begin{tabular}{ccc}
              \toprule
                $\CB$  & $\CH_0$  &  $\CH_1$  \\
                \hline 
                $1$  & $\BZ$  & $\BZ$ \\
                $-1$  & $\BZ$  & $\BZ+\tfrac{1}{2}$  \\
                \bottomrule
              \end{tabular}
              \vspace*{.3cm}
              
            \caption{$\BZ_2$ symmetry}
    \end{subtable}
    \hspace*{2.5cm}
    \begin{subtable}[b]{0.3\textwidth}
    \begin{tabular}{cccc}
              \toprule
                $\CB$  & $\CH_0$  &  $\CH_1$  & $\CH_2$ \\
                \hline 
                $1$  & $\BZ$  & $\BZ$  &  $\BZ$ \\
                $\omega$  & $\BZ$  & $\BZ+\tfrac{1}{3}$  &  $\BZ+\tfrac{2}{3}$  \\
                $\omega^2$ & $\BZ$  & $\BZ+\tfrac{2}{3}$  &  $\BZ+\tfrac{1}{3}$   \\\bottomrule
              \end{tabular}
              \caption{$\BZ_3$ symmetry}
    \end{subtable}
    \caption{The spin selection rules for non-anomalous $\BZ_2$ and $\BZ_3$ symmetries. Each row is labeled by the $b$-th twisted sector $\CH_b$ and the column is by the $\BZ_N$ charge $ \omega_N^a$ $(a\in\BZ_N)$ of the twisted sectors.}
    \label{tab:spin_selec}
\end{table}

\subsection{Orbifold and parafermionization by $\BZ_N$-gauging}\label{ss:orf_paraferm_ZN_gauging_general}

Let $Z_\CB[a,b]$ be the torus partition function of $\CH^a_{b}$, 
\begin{align}\label{PF_bosonic_ab}
    Z_\CB[a,b]
        :=
            \Tr_{\CH^a_{b}}\left[ q^{L_0 - \frac{c}{24}}\,\bar q^{\bar L_0 - \frac{\bar c}{24}}\right] \ .
\end{align}
It follows from \eqref{decomposition_hilbert_spaces} that this bosonic partition function is related to the torus partition function $\SCZ_\CB[\alpha, \beta]$ defined in \eqref{bosonic_PF} as\footnote{A similar relation has been introduced for $\BZ_2$-gauging in \cite{Gaiotto:2020iye}.}
\begin{align}\label{bosonic_PF_transformation}
    \SCZ_\CB[\alpha, \beta]
        =
            \sum_{a, b\in \BZ_N}\,\omega_N^{\alpha a}\,\delta_{\beta b}^{[N]}\,Z_\CB[a,b] \ ,
\end{align}
where $\delta_{a b}^{[N]}$ is the delta function mod $N$:  
\begin{align}
    \delta_{a b}^{[N]}
        :=
            \begin{cases}
                \,1 & a = b \pmod{N} \\
                \,0 & a \neq b \pmod{N} 
            \end{cases} \ .
\end{align}

For $1\le \rho < N$ coprime with $N$, one can define the orbifolded and parafermionized theories \cite{Thorngren:2021yso,Yao:2020dqx,Duan:2023ykn}, whose partition functions are given by
\begin{align}
    \SCZ_{\CO_\rho}[\alpha, \beta]
        &=
            \frac{1}{N}\,\sum_{\alpha', \beta'\in \BZ_N}\,\omega_N^{\rho(\beta\alpha' - \alpha\beta')}\,\SCZ_\CB[\alpha', \beta'] \ , \label{orbifold_PF}\\
    \SCZ_{\text{PF}_\rho}[\alpha, \beta]
        &=
            \frac{1}{N}\,\sum_{\alpha', \beta'\in \BZ_N}\,\omega_N^{\rho(\alpha + \alpha')(\beta+\beta')}\,\SCZ_\CB[\alpha', \beta']  \ .   \label{parafermion_PF}
\end{align}
Let $(\CH^{\CO_\rho})_{b}^{a}$ and $(\CH^{\text{PF}_\rho})_{b}^{a}$ be the Hilbert spaces of the $b$-th twisted sectors in the orbifolded and parafermionized theories, respectively, whose charges under the $\BZ_N$ symmetry operator $g$ are $a\in \BZ_N$.
Using the partition functions $Z_{\CO_\rho}[a,b]$ and $Z_{\text{PF}_\rho}[a,b]$ for $(\CH^{\CO_\rho})_{b}^{a}$ and $(\CH^{\text{PF}_\rho})_{b}^{a}$, the orbifolded and parafermionized partition functions can be expanded as
\begin{align}
    \SCZ_{\CO_\rho}[\alpha, \beta]
        &=
        \sum_{a,b\in \BZ_N}\,\omega_N^{\alpha a}\,\delta_{\beta b}^{[N]}\,Z_{\CO_\rho}[a,b] \ ,\\
    \SCZ_{\text{PF}_\rho}[\alpha, \beta]
        &=
        \sum_{a,b\in \BZ_N}\,\omega_N^{\alpha a}\,\delta_{\beta b}^{[N]}\,Z_{\text{PF}_\rho}[a,b] \ .
\end{align}
By inverting the first relation, we find
\begin{align}\label{orbfold_bosonic}
    \begin{aligned}
        Z_{\CO_\rho}[a,b] 
            &=
                \frac{1}{N}\,\sum_{\alpha,\, \beta\in\BZ_N} \omega_N^{-a\alpha}\,\delta_{b\beta}^{[N]}\,\SCZ_{\CO_\rho}[\alpha, \beta] \\
            &=
                Z_\CB\left[\,\langle -\rho\, b\rangle,\, \langle l\rangle\,\right] \ ,
    \end{aligned}
\end{align}
where we use \eqref{bosonic_PF_transformation} and \eqref{orbifold_PF}.
Here, $\langle x \rangle$ is the mod $N$ function, $\langle x \rangle\in \BZ_N$, and $l$ is the solution to the equation $a + \rho\,l = 0~\text{mod}~ N$, respectively.
Similarly, we find
\begin{align}\label{PF_bosonic}
    \begin{aligned}
        Z_{\text{PF}_\rho}[a,b]
            &=
                Z_\CB\left[\,\langle -a\rangle,\, \langle l'\rangle\,\right] \ ,
    \end{aligned}
\end{align}
where $l'$ is the solution to the equation $\rho(b + l') = a~\text{mod}~ N$.\par
It follows from Euler's theorem that coprime numbers $\rho,N$ satisfy
\begin{align}\label{rho_relation}
    \rho^{\varphi(N)}=1 \quad\text{mod}~ N \ ,
\end{align}
where $\varphi(N)$ is the Euler's totient function which is the number of positive integers up to $N$ that are coprime with $N$.\footnote{For prime $N$, Euler's theorem reduces to Fermat's little theorem:
\begin{align}
    \rho^{N-1}=1 \quad\text{mod}~ N \ .
\end{align}
}
By using this fact, the parameters $l,\,l'$ in \eqref{orbfold_bosonic} and \eqref{PF_bosonic} can be written as
\begin{align}
    l=-a\,\rho^{\varphi(N)-1}\ ,\qquad l'=a\,\rho^{\varphi(N)-1}-b \ .
\end{align}

To recapitulate the discussion so far, we have found that the Hilbert spaces in the bosonic theory are related to those in the orbifolded/parafermionized theories as follows:\footnote{While we will concentrate on the case with prime $N$ in the rest of this paper, the relations \eqref{correspondence_hilbert_spaces} hold for any $N\in \BZ_{\ge 2}$.}
\begin{align}\label{correspondence_hilbert_spaces}
    \begin{aligned}
    (\CH^{\CO_\rho})_{b}^{a}
        &=
            \CH_{\langle -a\,\rho^{\varphi(N)-1}\rangle}^{\langle -\rho\,b\rangle} \ , \\
    (\CH^{\text{PF}_\rho})_{b}^{a}
        &=
            \CH_{\langle a\,\rho^{\varphi(N)-1} - b\rangle}^{\langle -a\rangle} \ .
    \end{aligned}
\end{align}
Combining these relations with \eqref{spin_selection_rule_bosonic} and using \eqref{rho_relation}, we obtain the spin selection rules for the orbifolded and parafermionized theories:
\begin{align}\label{spin_selection_rule_orb_para}
    \begin{aligned}
        (\CH^{\CO_\rho})_{b}^{a}\, :
            & \quad
               & s 
                    &\in 
                        \BZ + \frac{a\,b}{N}\ , \\
        (\CH^{\text{PF}_\rho})_{b}^{a}\, :
            &\quad
                & s 
                    &\in 
                        \BZ + \frac{a\,b}{N} - \frac{a^2\,\rho^{\varphi(N)-1}}{N} \ .
    \end{aligned}
\end{align}
It follows that the spin of an operator in the untwisted sector $(\CH^{\CO_\rho})_{0}$ of the orbifolded theory is always integer while there are operators of fractional spins in the untwisted sector of the parafermionized theory $(\CH^{\text{PF}_\rho})_{0}$ as expected.

Let us illustrate the relations \eqref{correspondence_hilbert_spaces} for the case with $N=3$.
There are two choices of the parameter $\rho$ ($\rho=1$ and $\rho=2$) which are coprime to $N=3$, thus $\varphi(3) = 2$.
The Hilbert spaces in the twisted sectors of the orbifolded and parafermionized theories for the $\BZ_3$-gaugings can be read off from \eqref{correspondence_hilbert_spaces} as in table \ref{table:Z3-gauging}.
The two orbifolded theories $\CO_1$ and $\CO_2$ have the same untwisted sector $\CH_0$, but they are different in the twisted sectors $\CH_1, \CH_2$.
In general, the two orbifolded theories $\CO_\rho$ and $\CO_{\rho'}$ for the $\BZ_N$-gauging differ only in the twisted sectors.

\begin{table}
  \centering
      \begin{subtable}[t]{0.3\textwidth}
        \centering
              \begin{tabular}{cccc}
              \toprule
                $\CB$  & $\CH_0$  &  $\CH_1$  & $\CH_2$ \\
                \hline 
                $1$  & $\CH^0_{0}$  & $\CH^0_{1}$  &  $\CH^0_{2}$ \\
                $\omega$  & $\CH^1_{0}$  & $\CH^1_{1}$  &  $\CH^1_{2}$  \\
                $\omega^2$ & $\CH^2_{0}$  & $\CH^2_{1}$  &  $\CH^2_{2}$   \\\bottomrule
              \end{tabular}
        \caption{Bosonic CFT}
      \end{subtable}
      \quad 
      \begin{subtable}[t]{0.3\textwidth}
            \centering
                  \begin{tabular}{cccc}
                  \toprule
                    $\CO_1$  & $\CH_0$  &  $\CH_1$  & $\CH_2$ \\
                    \hline 
                    $1$  & $\CH^0_{0}$  & $\CH^2_{0}$  &  $\CH^1_{0}$ \\
                    $\omega$  & $\CH^0_{2}$  & $\CH^2_{2}$  &  $\CH^1_{2}$  \\
                    $\omega^2$ & $\CH^0_{1}$  & $\CH^2_{1}$  &  $\CH^1_{1}$   \\\bottomrule
                  \end{tabular}
            \caption{Orbifolded CFT ($\rho=1$)}
      \end{subtable}
      \quad
      \begin{subtable}[t]{0.3\textwidth}
            \centering
                  \begin{tabular}{cccc}
                  \toprule
                    $\CO_2$  & $\CH_0$  &  $\CH_1$  & $\CH_2$ \\
                    \hline 
                    $1$  & $\CH^0_{0}$  & $\CH^1_{0}$  &  $\CH^2_{0}$ \\
                    $\omega$  & $\CH^0_{1}$  & $\CH^1_{1}$  &  $\CH^2_{1}$  \\
                    $\omega^2$ & $\CH^0_{2}$  & $\CH^1_{2}$  &  $\CH^2_{2}$   \\\bottomrule
                  \end{tabular}
            \caption{Orbifolded CFT ($\rho=2$)}
      \end{subtable}
        \\

    \vspace*{1cm}
      
        \begin{subtable}[t]{0.4\textwidth}
            \centering
              \begin{tabular}{cccc}
              \toprule
                $\text{PF}_1$  & $\CH_0$  &  $\CH_1$  & $\CH_2$ \\
                \hline 
                $1$  & $\CH^0_{0}$  & $\CH^0_{2}$  &  $\CH^0_{1}$ \\
                $\omega$  & $\CH^2_{1}$  & $\CH^2_{0}$  &  $\CH^2_{2}$  \\
                $\omega^2$ & $\CH^1_{2}$  & $\CH^1_{1}$  &  $\CH^1_{0}$   \\\bottomrule
              \end{tabular}
            \caption{Parafermionized CFT ($\rho=1$)}
        \end{subtable}      
        \qquad
        \begin{subtable}[t]{0.4\textwidth}
            \centering
              \begin{tabular}{cccc}
              \toprule
                $\text{PF}_2$  & $\CH_0$  &  $\CH_1$  & $\CH_2$ \\
                \hline 
                $1$  & $\CH^0_{0}$  & $\CH^0_{2}$  &  $\CH^0_{1}$ \\
                $\omega$  & $\CH^2_{2}$  & $\CH^2_{1}$  &  $\CH^2_{0}$  \\
                $\omega^2$ & $\CH^1_{1}$  & $\CH^1_{0}$  &  $\CH^1_{2}$   \\\bottomrule
              \end{tabular}
            \caption{Parafermionized CFT ($\rho=2$)}
        \end{subtable}
        
  \vspace{0.5cm}
  \caption{The orbifolded and parafermionized theories of a bosonic CFT by the $\BZ_3$-gaugings.
  The row in each sub-table is labeled by the $b$-th twisted sector $\CH_b$ while the column is labeled by the $\BZ_3$ charge $\omega^a~(a\in \BZ_3)$.
  The two orbifolded theories $\CO_1$ and $\CO_2$ have the same untwisted sector $\CH_0$, but they are different in the twisted sectors $\CH_1, \CH_2$.
  }
  \label{table:Z3-gauging}
\end{table}

\section{$\BZ_N$-gauging of lattice CFTs}
\label{ss:ZN-gauging-lattice-CFTs}
The aim of this section is to formulate $\BZ_N$-gaugings in Narain CFTs, bosonic lattice CFTs whose momenta are characterized by even self-dual lattices $\Gamma$ \cite{Narain:1985jj,Narain:1986am}.
Section \ref{subsection;Narain CFTs and momentum lattices} reviews Narain CFTs with the same number of the left- and right-moving bosons.
We then move onto the discussion on general Narain/lattice CFTs with arbitrary numbers of the left- and right-moving bosons in section \ref{ss:lattice_CFT}.
We introduce new lattices $\Lambda$ related to the momentum lattices $\Gamma$ by an orthogonal transformation, which have the same structure as the Construction A lattices used extensively in section \ref{ss:codeCFT}.
In section \ref{ss:ZN-gauging_lattice_deform}, we describe the $\BZ_N$-gaugings of lattice CFTs as modifications of $\Lambda$ by a lattice vector $\chi$ that specifies a $\BZ_N$ symmetry.
Along the way, we derive the condition on $\chi$ for the $\BZ_N$ symmetry to be non-anomalous.

\subsection{Narain CFTs and momentum lattices}
\label{subsection;Narain CFTs and momentum lattices}
A Narain CFT is a two-dimensional field theory with $n$ compact bosons $X^i(t,\sigma)~ (i=1,\cdots,n)$, described by the following action \cite{Narain:1985jj,Narain:1986am}:
\begin{align}
    I
        =
        \dfrac{1}{4\pi\alpha'}\int \d t\int_{0}^{2\pi}\d\sigma\left[\,
        G_{ij}\,\left(\partial_tX^i\partial_tX^j-\partial_{\sigma}X^i\partial_{\sigma}X^j\right)
        -
        2B_{ij}\,\partial_tX^i\partial_{\sigma}X^j\,\right]\ ,
\end{align}
where $G_{ij}$ is the metric for a target space of the bosons and $B_{ij}$ is the anti-symmetric matrix called the B-field.
The target space of these bosons is compactified on a higher-dimensional torus with radius $R$: $X^i \sim X^i + 2\pi R$, thus there are winding modes satisfying
 \begin{align}
    X^i(t,\sigma)-X^i(t,\sigma+2\pi)=2\pi R\,w^i\ ,
\end{align}
where $w^i\in\BZ~ (i=1,\cdots,n)$ are the winding numbers.
This compactification makes the eigenvalues of the left- and right-moving momenta discretized as follows:
\begin{align}
    \begin{aligned}
         \hat{p}_{Li}
            &=
            \dfrac{m_i}{R} + \dfrac{R}{\alpha'}\,(B+G)_{ij}\,w^j\ ,\\
         \hat{p}_{Ri}
            &=
            \dfrac{m_i}{R} + \dfrac{R}{\alpha'}\,(B-G)_{ij}\,w^j\ ,
    \end{aligned}
\end{align}
where $m^i \in \BZ~\,(i=1,\cdots,n)$.
We also find it useful to introduce the dimensionless momenta:
\begin{align}
\label{dimensionless_momenta}
    \begin{aligned}
            p_{L\mu}
                &=
                e^{i}_{\mu}\left[\dfrac{m_i}{r}+\dfrac{r}{2}\,(B+G)_{ij}\,w^j\right] \ ,\\
            p_{R\mu}
                &=
                e^{i}_{\mu}\left[\dfrac{m_i}{r}+\dfrac{r}{2}\,(B-G)_{ij}\,w^j\right]\ ,
    \end{aligned}
\end{align}
where $e^{\mu}_{i}$ are vielbein satisfying $G_{ij}=e^{\mu}_{i}\,e^{\nu}_{j}\,\delta_{\mu\nu}$, $e^{i}_{\mu}$ their inverse, and $r:= R\sqrt{2/\alpha^{\prime}}$ is the dimensionless radius.
The left- and right-moving momenta form the momentum lattice $\Gamma$ whose lattice points are labeled by $(p_L,p_R)\in\BR^{n,n}$. For a Narain CFT to be modular invariant, the momentum lattice $\Gamma$ must be even, i.e., all the lattice vectors have even norms,
\begin{align}\label{even_lattice}
           ^{\forall}(p_L,p_R)\in\Gamma\ ,\qquad (p_L,p_R)\circledcirc_\Gamma (p_L,p_R)= p_L^2-p_R^2 \in 2\,\BZ\ ,
\end{align}
and also self-dual, $\Gamma^{*}=\Gamma$, where $\Gamma^{*}$ is the dual lattice of $\Gamma$ defined by
\begin{align}\label{dual_lattice}
           \Gamma^{*}=\left\{(p_L',p_R')\in \BR^{n,n}\ \bigg|\ ^{\forall}(p_L,p_R)\in\Gamma,\ (p_L,p_R) \circledcirc_\Gamma (p_L',p_R')\in\BZ\right\}\ .
\end{align}
Here, $\circledcirc_\Gamma$ stands for the Lorentzian inner product defined by
\begin{align}
    (p_L',p_R')\circledcirc_\Gamma(p_L,p_R)
        :=
            (p_L',p_R')\,\eta_\Gamma\,(p_L,p_R)^{T}\ ,
\end{align}
with the Lorentzian metric $\eta_\Gamma$ :
\begin{align}
\label{def:eta_tilde}
    \eta_\Gamma 
        =
        \left[
        \begin{array}{cc}
    	\,I_{n}~ & ~0 \\
            \,0~ & -I_n
        \end{array}
        \right] \ .
\end{align}
The momentum lattice $\Gamma$ is related to another lattice $\Lambda$ by the following coordinate transformation:
\begin{align}
\label{lambda-transformation}
    (\lambda_1,\lambda_2)=\left(\dfrac{p_L + p_R}{\sqrt{2}},\,\dfrac{p_L - p_R}{\sqrt{2}}\right)\in\Lambda\ ,\qquad (p_L,p_R)\in\Gamma\ .
\end{align}
In this basis, the Lorentzian inner product denoted by $\circledcirc_\Lambda$ is given from \eqref{def:eta_tilde} as
\begin{align}
\lambda,\lambda'\in\Lambda\ ,\qquad\lambda\circledcirc_\Lambda\lambda':=\lambda\,\eta_\Lambda\,\lambda^{'T} \ ,
\end{align}
where $\eta_\Lambda$ is the off-diagonal metric of the form:
\begin{align}
\label{def:eta_Lambda}
    \eta_\Lambda:=\left[
        \begin{array}{cc}
    	0~ & ~I_{n}~ \\
            I_{n}~ & ~0~
        \end{array}
        \right] \ .
\end{align}
Since the transformation \eqref{lambda-transformation} between $\Gamma$ and $\Lambda$ is orthogonal, an even self-dual momentum lattice $\Gamma$ with respect to $\eta_\Gamma$ yields an even self-dual lattice $\Lambda$ with respect to the metric $\eta_\Lambda$.
The generator matrix of $\Lambda$ corresponding to the dimensionless momenta \eqref{dimensionless_momenta} can be written as
\begin{align}
    M
        =
        \begin{bmatrix}
            \frac{\sqrt{2}}{r}\,\gamma^{-1}& \frac{r}{\sqrt{2}}\,B\\
            0&\frac{r}{\sqrt{2}}\,\gamma^{T}
        \end{bmatrix}\ ,
\end{align}
where $(\gamma)_{i\mu}:= e^{\mu}_{i}$.
Each lattice point of $\Lambda$ can be represented as a linear combination of the rows of the generator matrix $M$.
Conversely, any even self-dual lattice yields the momentum lattice of a modular invariant Narain CFT. 
Given a $2n$-dimensional even self-dual momentum lattice, we can calculate the torus partition function of the Narain CFT as 
\begin{align}\label{torus_PF_def}
    \begin{aligned}
        Z(\tau,\Bar{\tau})
            &=
            \Tr_\CH\left[q^{L_0-\frac{n}{24}}\,\Bar{q}^{\Bar{L}_0-\frac{n}{24}}\right]\\
            &=
            \dfrac{1}{|\eta(\tau)|^{2n}}\,\Theta_{\Gamma}(\tau,\Bar{\tau})\ ,
    \end{aligned}
\end{align}
where $\eta(\tau)$ is the Dedekind eta function,
\begin{align}
    \eta(\tau)  
        :=
            q^{\frac{1}{24}}\prod_{m=1}^{\infty}(1-q^m)\ ,
\end{align}
and $\Theta_{\Gamma}(\tau,\Bar{\tau})$ is the Siegel-Narain theta function,
\begin{align}
    \Theta_{\Gamma}(\tau,\Bar{\tau})
        :=
            \sum_{(p_L,p_R)\in\Gamma}q^{\frac{p_L^2}{2}}\Bar{q}^{\frac{p_R^2}{2}}\ .
\end{align}

\subsection{Lattice CFTs}
\label{ss:lattice_CFT}

In the previous subsection, we reviewed Narain CFTs with the same number of left- and right-moving bosons.
We saw that the spectrum of a Narain CFT is determined by the momentum lattice $\Gamma$, and the theory is bosonic and modular invariant if and only if $\Gamma$ is an even and self-dual lattice with respect to the Lorentzian metric \eqref{def:eta_tilde}.
This construction can be straightforwardly generalized to the case with $m$ left-moving and $n$ right-moving bosons as follows \cite{Narain:1986am,Narain:1985jj}.

Let $\Gamma$ be an $(m+n)$-dimensional lattice, and $\gamma = (p_L, p_R) \in \Gamma$ be a lattice point where $p_L$ and $p_R$ are $m$- and $n$-dimensional vectors, respectively.
We define the diagonal $(m,n)$ Lorentzian metric
\begin{align}\label{diagonal-Lorentzian-metric-mn}
    \eta^{(m,n)}_\Gamma
        :=
            \begin{bmatrix}
                I_m & 0 \\
                0 & - I_n
            \end{bmatrix} \ ,
\end{align}
which introduces the inner product $\circledcirc_\Gamma$ on $\Gamma$ as in the previous subsection\footnote{To avoid the clutter, we will use the same symbol $\circledcirc_\Gamma$ as the one for the case with $m=n$.}
\begin{align}
    p \circledcirc_\Gamma  p'
        :=
            (p_L,p_R)\,\eta_\Gamma^{(m,n)}\,(p_L',p_R')^{T}
        =
            p_L \cdot p_L' - p_R \cdot p_R'
            \ .
\end{align}
Evenness and dual lattices can also be defined in a similar manner to \eqref{even_lattice} and \eqref{dual_lattice}, respectively.

Given a lattice $\Gamma$, one can define a CFT with central charges $c = m$ and $\bar c =n$, whose primary states are given by $\ket{\,p_L, p_R\,}$ with $\gamma=(p_L, p_R) \in \Gamma$.
These states correspond to the primary operators of the form:
\begin{align}
    V_\gamma(z, \bar z) 
        = \,
            : e^{\i\, p_L\cdot X_L(z) + \i\,p_R\cdot X_R(\bar z)} : \ ,
\end{align}
where $X_L^a(z)~(a=1,\cdots,m)$ and $X_R^i(\bar z)~(i=1,\cdots,n)$ are compact bosons with the periodicity $(X_L, X_R) \sim (X_L, X_R) + 2\pi\gamma$ for any $\gamma\in \Gamma$.\footnote{In our convention, $X_L$ and $X_R$ satisfy $X_L^a(z)\,X_L^b(0) \sim - \delta^{ab} \log z$ and $X_R^i(\bar z)\,X_R^j(0) \sim - \delta^{ij} \log \bar z$, respectively.
We also do not keep track of the contribution of cocycle factors to vertex operators as they do not modify the following analysis.}
The OPE of a pair of the vertex operators $V_\gamma$ and $V_{\gamma'}$ corresponding to lattice vectors $\gamma=(p_L, p_R)$ and $\gamma'=(p_L', p_R')$ becomes
\begin{align}\label{OPE_vertex_operators}
     V_\gamma(z, \bar z) \,  V_{\gamma'}(0, 0) 
        \sim
            z^{p_L\cdot p_L'}\, \bar z^{p_R\cdot p_R'}\left[ V_{\gamma + \gamma'}(0, 0) + O(z, \bar z)\right] \ .
\end{align}
The operator $V_\gamma$ has the conformal dimension 
\begin{align}
    (h,\bar h) = \left( \frac{p_L^2}{2},\, \frac{p_R^2}{2}\right) \ ,
\end{align}
and has the spin 
\begin{align}
\label{eq:spin_vertex}
    s = h - \bar h = \frac{p_L^2 - p_R^2}{2} = \frac{\gamma^2}{2} \ ,
\end{align}
which implies that the operators are bosonic if $\Gamma$ is an even lattice.\footnote{Here, we use the short-hand notation; $\gamma^2:= \gamma\circledcirc_\Gamma \gamma$.}
On the other hand, the torus partition function of the CFT is given by
\begin{align}
    \begin{aligned}
        Z_\Gamma(\tau,\Bar{\tau})
            &=
            \Tr_\CH\left[q^{L_0-\frac{m}{24}}\,\Bar{q}^{\Bar{L}_0-\frac{n}{24}}\right]\\
            &=
            \dfrac{1}{\eta(\tau)^m\,\overline{\eta(\tau)^n}}\,\Theta_{\Gamma}(\tau,\Bar{\tau})\ .
    \end{aligned}
\end{align}
Then, the partition function for an even lattice $\Gamma$ transforms under the modular transformation as\footnote{
The Dedekind eta and Siegel-Narain functions transform under the modular $S$ transformation as
\begin{align}
    \eta\left( - \frac{1}{\tau}\right)
        &=
            (-\i\,\tau)^\frac{1}{2}\,\eta(\tau) \ , \\
    \Theta_{\Gamma}\left(- \frac{1}{\tau},- \frac{1}{\bar\tau}\right)
        &=
           \frac{ (-\i\,\tau)^\frac{m}{2}\,(\i\,\bar\tau)^\frac{n}{2}}{\text{vol}(\Gamma)}\,\Theta_{\Gamma^\ast}(\tau,\Bar{\tau}) \ ,
\end{align}
where $\text{vol}(\Gamma)$ is the volume of the unit cell in $\Gamma$, and the modular $T$ transformation as
\begin{align}
     \eta\left( \tau + 1 \right)
        &=
            e^\frac{\i\pi}{12}\,\eta(\tau) \ , \\
    \Theta_{\Gamma}\left(\tau + 1, \bar\tau + 1\right)
        &=
            \sum_{\gamma=(p_L,p_R)\in\Gamma}e^{\i\pi \gamma^2}q^{\frac{p_L^2}{2}}\Bar{q}^{\frac{p_R^2}{2}}
           \ .
\end{align}
}
\begin{align}
    Z_\Gamma\left(- \frac{1}{\tau},- \frac{1}{\bar\tau}\right)
        &=
            \frac{1}{\text{vol}(\Gamma)}\,Z_{\Gamma^\ast}(\tau,\Bar{\tau}) \ , \\
    Z_\Gamma(\tau + 1,\Bar{\tau} + 1)
        &=
             e^{- \frac{\i\pi (m-n)}{12}}\,Z_\Gamma(\tau,\Bar{\tau}) \ .
\end{align}
Hence one can construct a modular invariant bosonic CFT from an even self-dual lattice $\Gamma$ with respect to the Lorentzian metric with $m-n \in 24\,\BZ$.
It is known that there exist even self-dual lattices iff $m-n \in 8\,\BZ$ \cite{serre2012course}.
We will refer to bosonic CFTs constructed from even self-dual lattices with $m-n \in 8\,\BZ$ as \emph{lattice CFTs} to distinguish them from Narain CFTs with the same number of the left- and right-moving bosons.

For later convenience, we assume $m\ge n$ and introduce a new lattice $\Lambda$ by
\begin{align}
    \Lambda
        :=
            \left\{ \, \lambda \in \BR^{m,n}\, |\, \lambda = \gamma\,O \ , ~ \gamma\in \Gamma\,\right\} \ ,
\end{align}
where 
\begin{align}\label{orthogonal_matrix}
    O 
        :=
            \begin{bmatrix}
                I_{m-n} & 0 & 0 \\
                0 & ~\frac{I_n}{\sqrt{2}}~ & ~\frac{I_n}{\sqrt{2}}~ \\
                0 & ~\frac{I_n}{\sqrt{2}}~ & -\frac{I_n}{\sqrt{2}}~
            \end{bmatrix} \ .
\end{align}
We then introduce another Lorentzian metric $\eta^{(m,n)}_\Lambda$ on $\Lambda$ which is related to the diagonal Lorentzian metric by
\begin{align}\label{off-diagonal-Lorentzian-metric-mn}
    \eta^{(m,n)}_\Lambda
        :=
            O^T\, \eta^{(m,n)}_\Gamma\,O 
        =
            \begin{bmatrix}
                I_{m-n} & ~0~ & ~0~ \\
                0 & ~0~ & ~I_n~ \\
                0 & ~I_n~ & ~0~
            \end{bmatrix} \ .
\end{align}
Note that $O$ satisfies $O^T = O$ and $O^2 = I_{m+n}$.
A lattice vector $\lambda \in \Lambda$ can be written as $\lambda = (\lambda_0, \lambda_1, \lambda_2)$ where $\lambda_0, \lambda_1, \lambda_2$ are $(m-n)$-, $n$- and $n$-dimensional vectors, respectively, which are related to a momentum lattice vector $\gamma = (p_L, p_R) \in \Gamma$ as follows:
\begin{align}
    \begin{aligned}
    \label{eq:momentum_nonchiral}
        \lambda_0^i
            &:=
                p_L^i & \quad &(i=1,\cdots, m-n) \ , \\
        \lambda_1^a
            &:=
                \frac{1}{\sqrt{2}}\,(p_L^{m-n+a} + p_R^a) &\quad &(a=1,\cdots, n) \ , \\
        \lambda_2^a
            &:=
                \frac{1}{\sqrt{2}}\,(p_L^{m-n+a} - p_R^a) &\quad &(a=1,\cdots, n) \ .
    \end{aligned}        
\end{align}
With this definition, the inner product between a pair of lattice vectors $\lambda, \lambda' \in \Lambda$ becomes
\begin{align}
    \lambda \circledcirc_\Lambda \lambda'
        := 
            (\lambda_0, \lambda_1,\lambda_2)\,\eta_\Lambda^{(m,n)}\,(\lambda_0', \lambda_1',\lambda_2')^{T} 
        =
        \lambda_0\cdot \lambda_0' + \lambda_1\cdot \lambda_2' + \lambda_2\cdot \lambda_1'
        \ .
\end{align}
For a lattice vector $\lambda\in \Lambda$, we define the vertex operator 
\begin{align}
    U_\lambda(z, \bar z) 
        :=
            V_{\lambda\,O}(z, \bar z) \ .
\end{align}
Using \eqref{OPE_vertex_operators}, the OPE between a pair of the vertex operators $U_\lambda$ and $U_{\lambda'}$ for $\lambda, \lambda' \in \Lambda$ is read as
\begin{align}\label{OPE_vertex_operators_Lambda}
     U_\lambda(z, \bar z) \,  U_{\lambda'}(0, 0) 
        \sim
            z^{\lambda_0\cdot \lambda_0' + \frac{1}{2}(\lambda_1 + \lambda_2)\cdot (\lambda_1' + \lambda_2')}\, \bar z^{\frac{1}{2}(\lambda_1 - \lambda_2)\cdot (\lambda_1' - \lambda_2')}\left[ U_{\lambda + \lambda'}(0, 0) + O(z, \bar z)\right] \ .
\end{align}

Note that for a pair of lattice vectors $\lambda, \lambda' \in \Lambda$ corresponding to the momentum lattice vectors $p,p'\in\Gamma$, we have
\begin{align}\label{inner_product_equivalence}
    \lambda \circledcirc_\Lambda \lambda'
        =
            p \circledcirc_\Gamma p' \ .
\end{align}
This relation ensures that the lattice $\Lambda$ is even self-dual with respect to the off-diagonal Lorentzian metric $\circledcirc_\Lambda$ if and only if the momentum lattice $\Gamma$ is so with respect to the diagonal Lorentzian metric $\circledcirc_\Gamma$.

A few remarks are in order:
\begin{itemize}
    \item When $m=n$, lattice CFTs reduce to Narain CFTs whose momentum lattices are even self-dual as in the previous subsection.

    \item When $m=n+ 8\,k$ with $n> 0$ and $k\in \BZ_{> 0}$, any even self-dual lattices can be obtained from $k$ copies of $\Gamma_{E_8}$ times $n$ copies of a two-dimensional square lattice with the metric
    $\left[
        \begin{smallmatrix}
            0 & 1 \\ 
            1 & 0
        \end{smallmatrix}
     \right]
    $ by acting with an $\SO(n+8k, n)$ transformation \cite{Narain:1985jj}.
    The resulting theories are invariant under the modular $S$ transformation and invariant up to the phase $e^{\frac{2\pi\i\,(m-n)}{24}}$ under the modular $T$ transformation.
    In particular, they are completely modular invariant when $m - n\in 24\,\BZ_{> 0}$.

    \item When $m\in 8\,\BZ_{>0}$ and $n=0$, lattice CFTs are chiral CFTs or meromorphic CFTs constructed from even self-dual Euclidean lattices \cite{Dolan:1989vr,Dolan:1994st}.
    There is the unique lattice $\Gamma_{E_8}$ for $m=8$ (the $E_8$ lattice), two lattices $\Gamma_{E_8} \times \Gamma_{E_8}$ and $\Gamma_{D_{16}}$ (the $\text{Spin}(32)/\BZ_2$ lattice), and twenty-four lattices called Niemeier lattices for $m=24$.
\end{itemize}

\subsection{$\BZ_N$-gauging and lattice modifications}\label{ss:ZN-gauging_lattice_deform}

In this subsection, we will show that the $\BZ_N$-gauging of lattice CFTs can be described as modifications of the associated momentum lattices.
While the $\BZ_N$-gauging is possible for any $N\in \BZ_{\ge 2}$, we will only deal with the case with prime $N$, where the spin of an operator in the $b$-th twisted sector $\CH_b$ can be determined by the $\BZ_N$ charge through the relation \eqref{graid_twisted}.
Since the momentum lattice $\Gamma$ and the related lattice $\Lambda$ are equivalent under an orthogonal transformation, they have the same information about the spectrum of the CFTs.
Thus, we will concentrate on the $\BZ_N$-gauging on $\Lambda$, which will be relevant to the later sections.

Let $\Lambda$ be an $(m,n)$-dimensional even self-dual lattice with respect to the metric $\circledcirc_\Lambda$.
We choose a lattice vector $\chi\in\Lambda$ which satisfies
\begin{align}\label{characteristic_vector}
    \frac{a}{N}\,\chi \not \in \Lambda\qquad (a=1,\cdots, N-1)\ .
\end{align}
The vector $\chi$ specifies a $\BZ_N$ subgroup in the $\U(1)_L^m \times \U(1)_R^n$ symmetry that act as a shift symmetry on the left- and right-moving bosons in a lattice CFT as follows.
Consider the vertex operator $U_\frac{\chi}{N}(z, \bar z)$.
It follows from the relation \eqref{OPE_vertex_operators_Lambda} that encircling another vertex operator $U_\lambda(z,\bar z)$ for  around $U_\frac{\chi}{N}(0, 0)$ gives a phase:
\begin{align}\label{action_of_V_chi}
    U_\lambda(z, \bar z)\,U_\frac{\chi}{N}(0,0)
        \xrightarrow[z \to e^{2\i\pi} z]{}
        e^{\frac{2\i\pi}{N}\,\chi\circledcirc_\Lambda \lambda}\,U_\lambda(z, \bar z)\,U_\frac{\chi}{N}(0,0) \ .
\end{align}
Since $\Lambda$ is an integral lattice, we have $\chi\circledcirc_\Lambda \lambda \in \BZ$ and \eqref{action_of_V_chi} implies that $U_\frac{\chi}{N}$ induces the $\BZ_N$ action on the vertex operator $U_\lambda$, which has the charge $\chi  \circledcirc_\Lambda \lambda ~\text{mod}~N$.

With the $\BZ_N$ symmetry at hand, one can grade the lattice vectors in $\Lambda$ by their $\BZ_N$ charges as
\begin{align}
    \Lambda
        =
        \bigoplus_{a\in \BZ_N}\,\Lambda^{a}_{0} \ ,
\end{align}
where $\Lambda^{a}_{0}$ is the subsector with $\BZ_N$ charge $a$:
\begin{align}
    \Lambda^{a}_{0}
        =
        \left\{ \lambda \in \Lambda\;\bigg |\; \chi  \circledcirc_\Lambda \lambda = a \quad \text{mod}~ N\right\} \qquad (a=0,1, \cdots,N-1)\ .
\end{align}

Next, we introduce $N$ modified lattices $\Lambda_b ~(b=0, 1, \cdots, N-1)$ by shifting the lattice $\Lambda$ by the vector $\chi$:
\begin{align}
    \Lambda_b
        :=
            \Lambda + \frac{b}{N}\,\chi \qquad (b=0, 1, \cdots, N-1) \ .
\end{align}
The vectors in $\Lambda_b$ correspond to the operators in the $b$-th twisted sector.
Indeed, for a vertex operator $U_{\lambda_0 \in \Lambda}$ in the untwisted sector with unit charge under the $\BZ_N$ symmetry, i.e., $\chi\circledcirc_\Lambda \lambda_0 = 1 ~\text{mod}~N$, the OPE \eqref{OPE_vertex_operators_Lambda} implies
\begin{align}
    U_{\lambda_0}(z, \bar z)\,U_{\lambda_b}(0,0)
        \xrightarrow[z \to e^{2\i\pi} z]{}
        e^{ 2\i\pi\frac{b}{N}\,}\,U_{\lambda_0}(z, \bar z)\,U_{\lambda_b}(0,0) \ ,
\end{align}
for any vector $\lambda_b \in \Lambda_b$.
This means that $U_{\lambda_b}$ is the point operator attached to the end of the topological defect line $\hat \CL^b$ where $\hat\CL$ is the symmetry defect for the $\BZ_N$ symmetry.
Hence, $U_{\lambda_b}$ is the operator in the $b$-th twisted sector.
As the $\BZ_N$ symmetry acts also on the twisted sectors, we can  grade the vectors in $\Lambda_b$ by their $\BZ_N$ charges as
\begin{align}
    \Lambda_b
        =
        \bigoplus_{a\in \BZ_N}\,\Lambda^{a}_{b} \ ,
\end{align}
where $\Lambda^{a}_{b}$ is the subsector with $\BZ_N$ charge $a$ respecting the spin selection rule \eqref{spin_selection_rule_bosonic}:
\begin{align}\label{Lambda_a_b}
    \Lambda^{a}_{b}
        =
        \left\{ \lambda \in \Lambda_b\;\bigg |\: s = \dfrac{\lambda^2}{2} \in\BZ + \dfrac{a\,b}{N}\right\} \qquad (a=0,1, \cdots,N-1)\ .
\end{align}

Now, we examine the constraint on the vector $\chi$ from the non-anomalous condition \eqref{Non-anomalous-condition-ZN} for the $\BZ_N$ symmetry.
Given a vector $\mu \in \Lambda^a_{b}$, there exists $\lambda \in \Lambda$ such that $\mu = \lambda + \frac{b}{N}\,\chi$.
Then, the non-anomalous condition \eqref{Non-anomalous-condition-ZN} becomes
\begin{align}
    \begin{aligned}
        s
            &=
                \frac{\mu^2}{2} 
            =
                 \frac{1}{2}\left( \lambda + \frac{b}{N}\,\chi\right)^2 
            \in
                \frac{\BZ}{N} \ .
    \end{aligned}
\end{align}
This holds for any $b\in \BZ_N$ if and only if the following condition on $\chi$ is satisfied:
\begin{align}
\label{eq:non_anom}
    \chi^2 := \chi\circledcirc_\Lambda\chi \in 2N\,\BZ \ .
\end{align}

The charge $a$ subsector $\Lambda_b^a$ in the $b$-th twisted sector should be related to a subsector $\Lambda^{a'}_{0}$ with some charge $a'$ by shifting with $\frac{b}{N}\chi$:
\begin{align}
    \begin{aligned}
        \Lambda^a_{b}
            =
            \Lambda^{a'}_{0} + \frac{b}{N}\,\chi\ .
    \end{aligned}
\end{align}
Naively, one may expect $a' = a$, but they are different as we will show below.
Consider a state in the Hilbert space $\CH_b^a$ of the $b$-th twisted sector with charge $a$ which corresponds to a vector $\mu \in \Lambda^a_{b}$.
Suppose that there exists $\lambda \in \Lambda^{a'}_{0}$ such that $\mu = \lambda + \frac{b}{N}\,\chi$.
As we saw in section \ref{ss:twisted_sector}, the spin $s$ of the state is related to the charge $a$ by \eqref{graid_twisted}, which, in the present case, yields
\begin{align}
    a'
        =
            a-b\,m \ ,
\end{align}
where $m$ is an integer satisfying $\chi^2=2\,N\,m$.
Thus, we find
\begin{align}\label{Lambda_a_b_shift_relation}
     \Lambda^a_{b}
            =
            \Lambda^{a-b\,\ell}_{0} + \frac{b}{N}\,\chi \ ,
\end{align}
where $\ell \in \BZ_N$ is defined as follows:
\begin{align}\label{def_ell}
    \chi^2 = 2\,N(N\,\BZ + \ell)\ . 
\end{align}
In particular, for $N=2$, the twisted sectors $\Lambda^a_{b}~(b\neq 0)$ are given by
\begin{align}
    \Lambda^0_{1}
        =
            \begin{dcases}
                \Lambda^0_{0} + \frac{\chi}{2}  &  \quad (\chi^2 = 8\,\BZ) \\
                \Lambda^1_{0} + \frac{\chi}{2}  &  \quad (\chi^2 = 8\,\BZ + 4) 
            \end{dcases} \ , \qquad
     \Lambda^1_{1}
        =
            \begin{dcases}
                \Lambda^1_{0} + \frac{\chi}{2}  &  \quad (\chi^2 = 8\,\BZ) \\
                \Lambda^0_{0} + \frac{\chi}{2}  &  \quad (\chi^2 = 8\,\BZ + 4) 
            \end{dcases}  \ ,   
\end{align}
which reproduce the previous results in \cite{Kawabata:2023iss}.

The modifications of the momentum lattice $\Gamma$ follow from those of $\Lambda$ by acting with the orthogonal transformation.
Namely, the charge $a$ subsector $\Gamma_b^a$ in the $b$-th twisted sector of the momentum lattice is given by
\begin{align}
    \Gamma_b^a
        := 
            \left\{\, \gamma \in \BR^{m,n}\, \bigg|\, \gamma = \lambda\,O\ , ~ \lambda \in \Lambda_b^a \right\} \ .
\end{align}
Since the states in the Hilbert space $\CH_b^a$ are in one-to-one correspondence with the vectors in $\Gamma_b^a$, one can define the sets of vectors $(\Gamma^{\CO_\rho})_b^a$ and $(\Gamma^{\text{PF}_\rho})_b^a$ that describe the charge $a$ subsectors in the $b$-th twisted sector in the orbifolded and parafermionized theories based on the relations \eqref{correspondence_hilbert_spaces} as follows:
\begin{align}\label{orb_ferm_lattice_relations}
    \begin{aligned}
        (\Gamma^{\CO_\rho})_b^a
            &:=
                 \Gamma_{\langle -a\,\rho^{\varphi(N)-1}\rangle}^{\langle -\rho\,b\rangle} \ , \\
        (\Gamma^{\text{PF}_\rho})_b^a
            &:=
                \Gamma_{\langle a\,\rho^{\varphi(N)-1} - b\rangle}^{\langle -a\rangle} \ .
    \end{aligned}
\end{align}

In the lattice CFT, the torus partition function \eqref{PF_bosonic_ab} in the Hilbert space $\CH_b^a$ of the lattice CFT takes the following form:
\begin{align}\label{PF_lattice_theta}
        Z_\CB[a,b]
            &=
                \dfrac{1}{\eta(\tau)^m\,\overline{\eta(\tau)^n}}\,\Theta_{\Gamma_b^a}(\tau,\Bar{\tau})\ .
\end{align}
It follows from the relations \eqref{orb_ferm_lattice_relations} (or \eqref{orbfold_bosonic} and \eqref{PF_bosonic}) that the partition functions of the charge $a$ subsector in the $b$-th twisted sector for the orbifolded and parafermionized theories can be written as
\begin{align}\label{orb_paraferm_PF_lattice}
    \begin{aligned}
        Z_{\CO_\rho}[a,b]
            &=
                \dfrac{1}{\eta(\tau)^m\,\overline{\eta(\tau)^n}}\,\Theta_{\CO_\rho}[a,b]\ , \\
        Z_{\text{PF}_\rho}[a,b]
            &=
                \dfrac{1}{\eta(\tau)^m\,\overline{\eta(\tau)^n}}\,\Theta_{\text{PF}_\rho}[a,b]\ ,
    \end{aligned}
\end{align}
where $\Theta_{\CO_\rho}[a,b]$ and $\Theta_{\text{PF}_\rho}[a,b]$ are defined by
\begin{align}\label{orb_paraferm_lattice}
    \begin{aligned}
        \Theta_{\CO_\rho}[a,b]
            &:=
                \Theta_{(\Gamma^{\CO_\rho})_b^a}(\tau, \Bar \tau) \ , \\
        \Theta_{\text{PF}_\rho}[a,b]
            &:=
                \Theta_{(\Gamma^{\text{PF}_\rho})_b^a}(\tau, \Bar \tau) \ .
    \end{aligned}
\end{align}

\section{$\BZ_N$-gauging of code CFTs}\label{ss:codeCFT}

We apply our formulation for $\BZ_N$-gaugings of lattice CFTs in section \ref{ss:ZN-gauging-lattice-CFTs}
to code CFTs, lattice CFTs whose momentum lattices are built from error-correcting codes. 
In section \ref{ss:codeCFT_untwisted}, we review the well-established constructions of lattice CFTs from codes over a finite field $\BF_p$ through the Construction A \cite{Dymarsky:2020qom,Yahagi:2022idq,Kawabata:2022jxt} with a slight generalization to theories with unequal left and right central charges.
In section \ref{ss:Z_N_gauged_code_CFTs}, we exploit the results for $\BZ_N$-gaugings of lattice CFTs to construct the orbifolded and parafermionized theories of code CFTs.
We express the torus partition functions in terms of the weight enumerator polynomials of the associated codes, which we show simplify for specific choices of $\BZ_N$ symmetries.

\subsection{Code CFTs}\label{ss:codeCFT_untwisted}
In section \ref{ss:lattice_CFT}, we formulated the construction of lattice CFTs, i.e., bosonic modular invariant CFTs from Lorentzian even self-dual lattices.
Now, we turn to the subclass of lattice CFTs, called \emph{code CFTs}, whose momentum lattices are associated with codes.

\subsubsection{From codes to lattices}\label{ss:code_to_lattice}

A code is defined as a subspace of a finite-dimensional vector space over a finite field. Throughout this paper, we consider a finite field $\BF_p:= \BZ_p = \{0,1,2,\cdots,p-1\}$ with a prime $p$.\footnote{We will use $\BF_p$ and $\BZ_p$ for prime $p$ interchangeably throughout this paper.}
If a code $C$ is a $k$-dimensional subspace of $\BF_{p}^n$, then $C$ is referred to as an $[n,k]_p$ code. An $[n,k]_p$ code $C$ can be represented as
\begin{align}
    C=\left\{c\in\BF_{p}^{n}\,\Big|\,c=x\,G\ , ~x\in\BF_{p}^{k}\right\}\ ,
\end{align}
where $G$ is a $k\times n$ matrix called a generator matrix.

Given an $[m+n,k]_p$ code $C$, one can construct a lattice
$\Lambda(C)$ via the Construction A \cite{conway2013sphere}:
\begin{align}\label{ConstructionA}
    \Lambda(C)
        :=
            \left\{\, \frac{c + p\,v}{\sqrt{p}}\, \bigg|\, c\in C\ , ~ v \in \BZ^{m+n} \right\} \ .
\end{align}
We then introduce the momentum lattice $\Gamma(C)$ by performing the orthogonal transformation on the Construction A lattice $\Lambda(C)$ by the matrix $O$ defined in \eqref{orthogonal_matrix}:
\begin{align}
    \Gamma(C)
        :=
            \left\{\, \gamma = (p_L,p_R)\in \BR^{m,n}\, \bigg|\, \gamma = \lambda\,O\ , ~ \lambda \in \Lambda(C) \right\} \ .
\end{align}
We define the code CFT associated with a code $C$ as the lattice CFT whose momentum lattice is $\Gamma(C)$.

For the resulting code CFT being bosonic and modular invariant, the momentum lattice $\Gamma(C)$ must be even self-dual with respect to the Lorentzian metric $\eta_\Gamma^{(m,n)}$ given by \eqref{diagonal-Lorentzian-metric-mn} as discussed in section \ref{ss:lattice_CFT}. 
It follows from the relation \eqref{inner_product_equivalence} that the Construction A lattice $\Lambda(C)$ must also be even self-dual with respect to the off-diagonal Lorentzian metric $\eta_\Lambda^{(m,n)}$.
This condition provides us some constraints on the code $C$ as we will see soon.

Having the construction of Lorentzian self-dual lattices in mind, we define the dual code $C^{\perp}$ with the metric $\eta_\Lambda^{(m,n)}$ by
\begin{align}
    C^\perp =  \left\{\, c'\in \BF_p^{m + n}\,\big|\,c'\circledcirc_\Lambda c= 0~\text{mod}~p\,, \forall c \in C\,\right\} \ ,
\end{align}
where we use the same symbol $\circledcirc_\Lambda$ for the codes as the one for the lattice.
The code $C$ is called self-orthogonal if $C\subset C^\perp$, and self-dual if $C=C^\perp$.
With this definition, it is easy to show that the Construction A lattice of the dual code $C^\perp$ is the dual lattice of $\Lambda(C)$, i.e., $\Lambda^\ast(C) = \Lambda(C^\perp)$.
In particular, the Construction A lattice $\Lambda(C)$ is self-dual with respect to the metric $\circledcirc_\Lambda$ if and only if the code $C$ is self-dual.
Furthermore, by taking into account the evenness condition, one can characterize a class of codes $C$ that generate even self-dual lattices $\Lambda(C)$ as follows:
\begin{itemize}
    \item For $m>n$, the Construction A lattice $\Lambda(C)$ is an even self-dual lattice with respect to the metric $\circledcirc_\Lambda$ if and only if $C$ is a doubly even self-dual $[m+n, \frac{m+n}{2}]_2$ code, where doubly even codes are binary $(p=2)$ codes satisfying
    \begin{align}
        ^{\forall}c= (c_0,c_1,c_2)\in C\ ,\quad c\circledcirc_\Lambda c =  c_0^2 + 2c_1\cdot c_2\in 4\BZ \ .
    \end{align}

    \item For $m=n$, the Construction A lattice $\Lambda(C)$ is an even self-dual lattice with respect to the metric $\circledcirc_\Lambda$ if and only if one of the followings holds \cite{Yahagi:2022idq}:    
        \begin{itemize}
            \item $p$ is an odd prime and $C$ is a self-dual $[2n, n]_p$ code with respect to the metric $\circledcirc_\Lambda$.
            \item $p=2$ and the code $C$ is a doubly even self-dual $[2n, n]_2$ code with respect to the metric $\circledcirc_\Lambda$.
        \end{itemize}
\end{itemize}

Using the generator matrix $G$, the self-duality of an $[m+n, \frac{m+n}{2}]_p$ code $C$ with prime $p$ can be written as
\begin{align}\label{code_self-duality}
        \text{(self-duality)}&  &
            G\,\eta_\Lambda^{(m,n)}\,G^T &= 0 \quad \pmod{p} \ .
\end{align}
For $m=n$, this condition also ensures that $C$ is even.
On the other hand, the doubly-evenness condition for $C$ with $m>n$ and $p=2$ becomes
\begin{align}\label{code_doubly-even}
    \text{(doubly-evenness)}&  & 
            \text{diag}\left(G\,\eta_\Lambda^{(m,n)}\,G^T\right) &= 0 \quad \pmod{4} \ .
\end{align}
In section \ref{ss:Examples}, we will provide concrete examples of codes that satisfy these conditions and construct the code CFTs as well as their $\BZ_N$-gaugings.

\subsubsection{Torus partition functions}

Now we consider the torus partition function of the code CFT. 
To this end, we define the complete weight enumerator polynomial of an $[m+n, k]_p$ code $C$ by
\begin{align}\label{CWE}
    W_{C}(\{ x_{\alpha}\}, \{x_{ab}\})
        :=
            \sum_{c\,\in\,C}\,\prod_{\alpha\in\BF_p} x_{\alpha}^{\text{wt}_{\alpha}(c_0)}\prod_{(a,b)\,\in\,\BF_p\times\BF_p }\, x_{ab}^{\text{wt}_{ab}\left( (c_1, c_2)\right)}\ ,
\end{align}
where for each codeword $c = (c_0, c_1, c_2)\in C$ with $c_0\in \BF_p^{m-n}$ and $c_1, c_2\in \BF_p^n$, the compositions $\text{wt}_{\alpha}(c_0)$ and $\text{wt}_{a b}\left( (c_1, c_2)\right)$ are defined by
\begin{align}
    \begin{aligned}
        \text{wt}_{\alpha}(c_0)
            &:=
                |\,\{i\in\{1,\cdots,m-n\}\,|\,(c_0)_i= \alpha\}\,| \ , \\
        \text{wt}_{ab}\left( (c_1, c_2)\right)
            &:=
                |\,\{i\in\{1,\cdots,n\}\,|\,\left( (c_1)_i,(c_2)_i \right)=(a, b)\}\,| \ .
    \end{aligned}
\end{align}
Using the complete weight enumerator polynomial, the partition function can be written as
\begin{align}
\label{partition_enumerator_relation}
    Z_\CB(\tau,\Bar{\tau})
        =
        \Tr_\CH\left[q^{L_0-\frac{m}{24}}\,\Bar{q}^{\Bar{L}_0-\frac{n}{24}}\right]
        =
        \dfrac{1}{\eta(\tau)^{m}\,\overline{\eta(\tau)^n}}\,W_{C}(\{\phi_{\alpha}\}, \{\psi_{ab}\})\ ,
\end{align}
where $\CH$ is the Hilbert space in the untwisted sector.
The arguments of the complete weight enumerator polynomial $\phi_\alpha$ and $\psi_{ab}$ are given by
\begin{align}\label{psi+}
    \begin{aligned}
    \phi_{\alpha}(\tau)
        &:=
             \Theta_{\alpha,\frac{p}{2}}(\tau) \ , \\
     \psi_{ab}(\tau,\Bar{\tau})
        &:=
            \Theta_{a+b,\,p}(\tau)\,\Bar{\Theta}_{a-b,\,p}(\Bar{\tau})+\Theta_{a+b-p,\,p}(\tau)\,\Bar{\Theta}_{a-b-p,\,p}(\Bar{\tau}) \ ,
    \end{aligned}
\end{align}
where $\Theta_{l,k}(\tau)$ is the theta function:
\begin{align}
    \Theta_{l,k}(\tau)
        =
            \sum_{v\in\BZ}\,q^{k\left(v +\frac{l}{2k}\right)^2}\ .
\end{align}
The functions $\phi_{\alpha}$ and $\psi_{ab}$ are invariant under the shifts of the indices $\alpha$ and $a, b$ by $p$ due to the property of the theta function, 
\begin{align}\label{theta_property}
    \Theta_{l,k}(\tau) = \Theta_{l+2k,\,k}(\tau) = \Theta_{-l,\,k}(\tau)\ .
\end{align}
This invariance is in accordance with the restriction of their values to $\BF_p$ in \eqref{CWE}.

We also note that the torus partition function \eqref{partition_enumerator_relation} incorporates both the result of \cite{Dymarsky:2020qom,Kawabata:2022jxt} for $m=n$ and that of \cite{Dolan:1994st,Gaiotto:2018ypj,Kawabata:2023nlt} for $n=0$ in a natural manner as expected.

\subsection{$\BZ_N$-gauged code CFTs}
\label{ss:Z_N_gauged_code_CFTs}
We will consider modular-invariant bosonic code CFTs with a non-anomalous $\BZ_N$ symmetry and perform their $\BZ_N$-gaugings by applying the methods given in section \ref{ss:ZN-gauging-lattice-CFTs}.

Let $C$ be an $[m+n,\frac{m+n}{2}]_p$ code and suppose there exists a vector $\chi$ in the Construction A lattice $\Lambda(C)$ which corresponds to a non-anomalous $\BZ_N$ symmetry.
Then, we define $\Lambda^a_b(C)$ by \eqref{Lambda_a_b} with $\Lambda$ replaced by $\Lambda(C)$.
We also introduce $\Gamma^a_b(C)$ by
\begin{align}
    \begin{aligned}
        \Gamma^a_b(C)
            :=
            \left\{\gamma=(p_L,p_R)\in\BR^{m,n}\,\Big|\, \gamma=\lambda\, O\ , ~ \lambda\in\Lambda^a_b(C)\right\} \ .
    \end{aligned}
\end{align}
A vector $\gamma$ in $\Gamma^a_b(C)$ can be seen as a momentum of a primary state in the $b$-th twisted sector $\CH^a_b$ of the code CFT with $\BZ_N$ charge $a$.
Thus, it follows from \eqref{PF_lattice_theta} that the partition function of the sector $\CH^a_{b}$ is given by
\begin{align}\label{PF_a_b}
    Z_\CB[a,b]
        =
            \dfrac{1}{\eta(\tau)^m\,\overline{\eta(\tau)}^n}\,\Theta_{\Gamma^a_{b}(C)}(\tau,\Bar{\tau}) \ .
\end{align}

Now, we will represent the partition function \eqref{PF_a_b} in terms of the weight enumerator polynomial as we did in section \ref{ss:codeCFT_untwisted} for the untwisted case.
Let us choose a lattice vector $\lambda$ in $\Lambda^a_{b}(C)$.
It follows from the relation \eqref{Lambda_a_b_shift_relation} that $\lambda$ can be written by using a vector $\mu\in\Lambda^{a-b\ell}_{0}(C)$ with $\ell$ defined by \eqref{def_ell} as
\begin{align}
    \lambda 
        =
            \mu + \frac{b}{N}\,\chi \ .
\end{align}
Since $\chi$ is a vector in $\Lambda(C)$, it takes the following form:
\begin{align}
\label{choice_of_chi}
    \chi
        =
            \left( \frac{\gamma_0 + p\,\nu_0}{\sqrt{p}}\ ,\, \frac{\gamma_1 + p\,\nu_1}{\sqrt{p}}\ ,  \, \frac{\gamma_2 + p\,\nu_2}{\sqrt{p}}\right) \ ,
\end{align}
where $(\gamma_0, \gamma_1, \gamma_2) \in C$ and $\nu_0\in \BZ^{m-n}$, $\nu_1, \nu_2 \in \BZ^n$.
Similarly, we represent $\mu$ as 
\begin{align}
    \mu
        =
            \left( \frac{\alpha_0 + p\,k_0}{\sqrt{p}}\ , \, \frac{\alpha_1 + p\,k_1}{\sqrt{p}}\ ,\,  \frac{\alpha_2 + p\,k_2}{\sqrt{p}}\right) \ ,
\end{align}
where $(\alpha_0, \alpha_1, \alpha_2) \in \BF_p^{m+n}$ and $k_0\in \BZ^{m-n}$, $k_1, k_2 \in \BZ^n$, respectively.
To ensure that $\mu$ is a vector in $\Lambda^{a-b\ell}_{0}(C)$, we must impose the condition
\begin{align}\label{grading_condition}
    \mu\circledcirc_\Lambda \chi = a - b\,\ell \quad \text{mod}~N \ .
\end{align}
Taking into account this condition and using the momentum $(p_L, p_R) = \lambda\,O \in \Gamma_b^a(C)$, a straightforward calculation shows that the theta function can be written as 
\begin{align}
\label{lattice_theta_of_each_sector}
    \begin{aligned}
        \Theta_{\Gamma^a_{b}(C)}(\tau,\Bar{\tau})
            &=
                \sum_{(\alpha_0,\alpha_1,\alpha_2)\in C}\;\sum_{\substack{k_0\in\BZ^{m-n}\\k_1,k_2\in\BZ^n}}\delta_{\mu\circledcirc_\Lambda\chi,\, a-b\ell}^{[N]}\,q^{\frac{p_L^2}{2}}\Bar{q}^{\frac{p_R^2}{2}}\\
            &=
                \dfrac{1}{N}\sum_{\hat a\in\BZ_N}\omega_N^{-\hat a (a-b\ell)}\,W_C^{\text{full}}\left( \left\{\phi_{\alpha_{0}^i}^{(\chi_{0}^i)}(\hat a, b)\right\},\left\{ \psi_{\alpha_{1}^i\alpha_{2}^i}^{(\chi_{1}^i,\chi_{2}^i)}(\hat a, b)\right\}\right) \ ,
    \end{aligned}
\end{align}
where we denote the $i$-th component of a vector $v$ as $v^i$ and $W_C^{\text{full}}$ is the full enumerator polynomial of an $[m+n, k]$ code $C$:
\begin{align}
    W_C^{\text{full}}\left( \left\{ x_{\alpha_{0}^i}^{(i)}\right\},\left\{ x_{\alpha_{1}^i\alpha_{2}^i}^{(i)}\right\}\right)
        :=
            \sum_{(\alpha_0,\alpha_1,\alpha_2)\in C}\,
            \left(\prod_{i=1}^{m-n} x_{\alpha_{0}^i}^{(i)}\right)
            \,
            \left(\prod_{i=1}^n x_{\alpha_{1}^i\alpha_{2}^i}^{(i)}\right)  \ .
\end{align}
The functions $\phi_{\alpha_{0}^i}^{(\chi_{0}^i)}(\hat a, b)$, $\psi_{\alpha_{1}^i\alpha_{2}^i}^{(\chi_{1}^i,\chi_{2}^i)}(\hat a, b)$ are defined by
\begin{align}
\begin{aligned}
    \phi_{\alpha_{0}^i}^{(\chi_{0}^i)}(\hat a, b)
        &:=
            \omega_{N}^{\frac{\hat a}{p}\left[ \alpha_{0}^i (\gamma_0 + p\,\nu_0)^i \right]}
            \sum_{k_0\in\BZ}
                \omega_{N}^{\hat a\left[ k_0 (\gamma_0 + p\,\nu_0)^i \right]}\,q^{\frac{(p_{L}^i)^2}{2}}\ ,\\
    \psi_{\alpha_{1}^i\alpha_{2}^i}^{(\chi_{1}^i,\chi_{2}^i)}(\hat a, b)
        &:=
            \omega_{N}^{\frac{\hat a}{p}\left[ \alpha_{1}^i (\gamma_2 + p\,\nu_2)^i + \alpha_{2}^i (\gamma_1 + p\,\nu_1)^i\right]}
            \sum_{k_1,k_2\in\BZ}
                \omega_{N}^{\hat a\left[ k_1 (\gamma_2 + p\,\nu_2)_i + k_2 (\gamma_1 + p\,\nu_1)_i\right]}\,q^{\frac{(p_{L}^{m-n+i})^2}{2}}\Bar{q}^{\frac{(p_{R}^i)^2}{2}}\ .
\end{aligned}
\end{align}

Let us choose a specific vector $\chi$ to simplify the full enumerator polynomial in the torus partition function. We consider the case in which the vector $\chi$ is given by
\begin{align}\label{characteristic_vec_non_chiral}
    \chi
        =
            \sqrt{p}\,\left(\,l_0\,{\bf 1}_{m-n},\,l_1\,{\bf 1}_{n},\,l_2\,{\bf 1}_{n}\right)\ , \qquad l_i\in\BZ ~~ (i=0,1,2)\ .
\end{align}
This vector can be obtained by choosing $(\gamma_0,\gamma_1, \gamma_2)=(\bf{0},\bf{0},\bf{0})$ and $(\nu_0,\nu_1, \nu_2)=\left(\,l_0\,{\bf 1}_{n},\, l_1\,{\bf 1}_{n},\, l_2\,{\bf 1}_{n}\right)$ in \eqref{choice_of_chi}.
Consider shifting $l_i~(i=0,1,2)$ by an integer multiple of $N$ as
\begin{align}
     l_i \to l_i + \kappa_i\, N \,,\qquad \kappa_i\in\BZ\ .
\end{align}
Then, the lattice vector \eqref{characteristic_vec_non_chiral} transforms under the shift as
\begin{align}
\begin{aligned}
    \chi &\to
        \chi+N\kappa \ , \qquad
        \kappa := \sqrt{p}\,(\,\kappa_0\,{\bf 1}_{m-n},\, \kappa_1\,{\bf 1}_{n},\, \kappa_2\,{\bf 1}_{n}) \ .
\end{aligned}
\end{align}
Note that $\kappa$ is included in $\Lambda(C)$ by choosing $(\gamma_0, \gamma_1, \gamma_2)=({\bf 0}_{m-n},{\bf 0}_n,{\bf 0}_n)$ and $(\nu_0, \nu_1, \nu_2)=(\,\kappa_0\,{\bf 1}_{m-n},\, \kappa_1\,{\bf 1}_{n},\,  \kappa_2\,{\bf 1}_{n}) $ in \eqref{choice_of_chi}.
We can show that this $N$ shift does not change the structure of each sector $\Lambda^a_b(C)$.
Thus, we only need to consider the case for $l_i\in\BZ_N~(i=0,1,2)$.
With this choice of $\chi$, the theta function \eqref{lattice_theta_of_each_sector} reduces to the following form:
\begin{align}
\label{lattice_theta_all_one}
    \begin{aligned}
        \Theta_{\Gamma^a_{b}(C)}(\tau,\Bar{\tau})                
                &=\dfrac{1}{N}\sum_{\hat a\in\BZ_N}\omega_N^{-\hat a (a-b\ell)}\,W_C\Big( \left\{\phi_{\alpha_{0}}(\hat a, b)\right\},\left\{ \psi_{\alpha_{1}\alpha_{2}}(\hat a, b)\right\}\Big) \ ,
                
    \end{aligned}
\end{align}
where $W_C$ is the complete weight enumerator polynomial defined by \eqref{CWE} and $\phi_{\alpha_{0}}(\hat a, b)$, $\psi_{\alpha_{1}\alpha_{2}}(\hat a, b)$ are defined by
\begin{align}
    \begin{aligned}
        \phi_{\alpha_{0}}(\hat a, b) &:= \omega_{N}^{\hat{a}l_0\alpha_0}\sum_{r\in\BZ_N}\omega_{N}^{\hat{a}l_0 pr}\,\Theta_{N(\alpha_0+pr)+l_0bp,\frac{pN^2}{2}}(\tau)\ ,\\
        \psi_{\alpha_{1}\alpha_{2}}(\hat a, b) 
            &:= 
                \omega_N^{\hat{a}(l_2\alpha_{1}+l_1\alpha_2)}
                     \left( 
                        \Phi_{+1, 0}(\tau)\,\bar \Phi_{-1, 0}(\bar\tau) 
                        +
                        \omega_N^{\hat{a}pl_2}\,   
                        \Phi_{+1, 1}(\tau)\,\bar \Phi_{-1, 1}(\bar\tau) 
                    \right) \ ,
    \end{aligned}
\end{align}
and
\begin{align}
    \Phi_{s, t}(\tau)
        :=
            \sum_{r\in\BZ_N}\omega_N^{(l_1 + sl_2)\hat{a}pr}\Theta_{N(\alpha_1 + s\alpha_2 + p(2r+t))+(l_1 + sl_2)bp,\,pN^2}(\tau) \ , \qquad (s= \pm 1, t = 0, 1) \ .
\end{align}

For the vector \eqref{characteristic_vec_non_chiral} to generate a non-trivial $\BZ_N$ symmetry, it has to satisfy the condition \eqref{characteristic_vector}.
Moreover, for the $\BZ_N$ symmetry to be non-anomalous, it must also satisfy the non-anomalous condition \eqref{eq:non_anom}, whose explicit form is given by
\begin{align}\label{non_anom_asymmetric}
    \begin{aligned}
        \chi^2
            =
                p\left[l_0^2\,(m-n)+2\,l_1\,l_2 \,n\right]\in 2N\,\BZ\ .
    \end{aligned}
\end{align}
When these two conditions are met, one can define the orbifolded and parafermionized theories of the non-chiral code CFT by performing the $\BZ_N$-gaugings.
The torus partition functions for the resulting orbifolded and parafermionized theories can be calculated from \eqref{orb_paraferm_PF_lattice} and \eqref{orb_paraferm_lattice} with the theta function \eqref{lattice_theta_all_one}.
We will examine these conditions and the $\BZ_N$-gaugings for a few concrete examples in the subsequent sections.

\section{Examples of $\BZ_N$-gaugings}\label{ss:Examples}
In this section, we demonstrate our formulation for $\BZ_N$-gaugings of code CFTs by several concrete examples.
In section \ref{ss:non-chiral-examples}, we consider non-chiral code CFTs with the same number of left and right central charges.
We construct the orbifolded/parafermionized theories of code CFTs with central charges $c=1$ and $c=2$ by $\BZ_3$-gaugings and find a few examples of code CFTs that are self-dual under $\BZ_3$-orbifolding.
In section \ref{ss:chiral-examples}, we consider $\BZ_3$-gaugings of chiral code CFTs with $c=24$ built from nine binary doubly even self-dual codes of length $24$ and obtain their orbifolded and parafermionized theories.
Section \ref{ss:asymmetric-examples} deals with code CFTs with asymmetric central charges, $c = m$ and $\bar c = n$.
We introduce CSS-like codes, which consist of three classical codes subject to certain relations.
The resulting code CFTs are shown to factorize into the chiral and non-chiral parts.
Inspecting the non-anomalous condition, we find a class of $\BZ_N$ symmetries which act non-anomalously on the whole theories while acting anomalously when restricted to the chiral and non-chiral sectors.
As an illustration, we examine several examples of CSS-like code CFTs with such $\BZ_3$ symmetries and calculate the torus partition functions of their orbifolded and parafermionized theories.

\subsection{Non-chiral code CFTs}\label{ss:non-chiral-examples}

In this subsection, we focus on code CFTs with $m=n$, which are called \emph{non-chiral code CFTs}.
The metric on a code is given by \eqref{def:eta_Lambda} and we denote the codewords by $(c_1,c_2)\in\BF_p^{n}\times\BF_p^{n}$ and lattice vectors as $(\lambda_1,\lambda_2)\in\BR^{n}\times\BR^{n}$.

\subsubsection{Torus partition functions}

We consider the torus partition functions of a non-chiral code CFT and their $\BZ_N$-gaugings.
By setting $m=n$ in \eqref{partition_enumerator_relation}, the partition function becomes
\begin{align}
    Z_\CB(\tau,\Bar{\tau})
        =
        \dfrac{1}{|\eta(\tau)|^{2n}}\,W_{C}(\{\psi_{\alpha_1\alpha_2}\})\ ,
\end{align}
where the complete weight enumerator polynomial \eqref{CWE} reduces to
\begin{align}
     W_{C}(\{x_{\alpha_1\alpha_2}\})
        :=
            \sum_{(c_1,c_2)\,\in\,C}\,\prod_{(\alpha_1,\alpha_2)\,\in\,\BF_p\times\BF_p }\, x_{\alpha_1\alpha_2}^{\text{wt}_{\alpha_1\alpha_2}\left( (c_1, c_2)\right)}\ .
\end{align}
Next, we perform the $\BZ_N$-gauging for the non-chiral code CFT in the same manner as in section \ref{ss:Z_N_gauged_code_CFTs}.
Let us specify the $\BZ_N$ symmetry by choosing the vector $\chi$ of the form \eqref{characteristic_vec_non_chiral} with $m=n$:
\begin{align}
\label{non_chiral_vector_chi}
     \chi=\sqrt{p}\,(\,l_1\,{\bf 1}_{n},\, l_2\,{\bf 1}_{n})\ ,\qquad l_1,l_2\in\BZ_N\ .
\end{align}
With this choice, the non-anomalous condition \eqref{non_anom_asymmetric} becomes
\begin{align}
\label{non_anomalous}
    \chi^2
        =
            2\,p\,n\,l_1\,l_2\in 2N\,\BZ\,,
\end{align}
and the condition \eqref{characteristic_vector} becomes
\begin{align}
\label{all_one_condition}
    \dfrac{a}{N}\,\chi
        =
            \dfrac{a}{N}\sqrt{p}\,(\,l_1\,{\bf 1}_{n}, \, l_2\,{\bf 1}_{n})\notin \Lambda(C)\qquad \text{for}~~ a= 1, \cdots, N-1 \ .
\end{align}
Since both $p$ and $N$ are prime in our consideration, we will deal with the cases with $p\neq N$ and $p= N$ separately below.

\paragraph{Case with $p\neq N$}
When $p\neq N$, the conditions \eqref{non_anomalous} and \eqref{all_one_condition} are simultaneously satisfied if and only if one of the following holds:
\begin{itemize}
    \item $l_1\,l_2 = 0$ and $(l_1, l_2)\neq (0,0)$\ .
    \item $l_1\,l_2\neq 0$ and $n\in N\BZ$\ .
\end{itemize}
The lattice theta function of each sector can be read off from \eqref{lattice_theta_all_one} as
\begin{align}
\label{lattice_theta_non_chiral}
    \begin{aligned}
         \Theta_{\Gamma^a_{b}(C)}(\tau,\Bar{\tau})
            =
                \dfrac{1}{N}\sum_{\hat{a}\in\BZ_N}\omega_N^{-\hat{a}(a-b\ell)}\, W_{C}\left(\{\psi_{\alpha_1\alpha_2}(\hat a,b)\}\right)\ .
    \end{aligned}
\end{align}
Using \eqref{def_ell} and \eqref{non_anomalous}, we obtain another expression of $\Theta_{\Gamma^a_{b}(C)}(\tau,\Bar{\tau})$:
\begin{align}
    \begin{aligned}
         \Theta_{\Gamma^a_{b}(C)}(\tau,\Bar{\tau})
            &=
                \dfrac{1}{N}\sum_{\hat{a}\in\BZ_N}\omega_N^{-\hat{a}a} \, W_{C}\left(\{\Tilde{\psi}_{\alpha_1\alpha_2}(\hat a,b)\}\right)\ ,
    \end{aligned}
\end{align}
where $\Tilde{\psi}_{\alpha_1\alpha_2}(\hat a,b) = \omega_N^{\,\frac{\hat{a}bpl_1l_2}{N}}\,\psi_{\alpha_1\alpha_2}(\hat a,b)$.
For $N=2$, it takes a simple form as
\begin{align}
    \begin{aligned}
        \Theta_{\Gamma^a_{b}(C)}(\tau,\Bar{\tau})
            &=
                \dfrac{1}{2}\bigg[ \, W_{C}\left(\{\Tilde{\psi}_{\alpha_1\alpha_2}(0,b)\}\right)+(-1)^{a}\,W_{C}\left(\{\Tilde{\psi}_{\alpha_1\alpha_2}(1,b)\}\right)\bigg]\ ,
    \end{aligned}
\end{align}
with
\begin{align}
    \begin{aligned}
         \Tilde{\psi}_{\alpha_1\alpha_2}(0,b)
            &=
                \Theta_{\alpha_1+\alpha_2+bp,\,p}(\tau)\, \Bar{\Theta}_{\alpha_1-\alpha_2,\,p}(\Bar{\tau})
                +
                \Theta_{\alpha_1+\alpha_2+(b+1)p,\,p}(\tau)\,\Bar{\Theta}_{\alpha_1-\alpha_2+p,\,p}(\Bar{\tau}) \ ,\\

         \Tilde{\psi}_{\alpha_1\alpha_2}(1,b)
            &=
                e^{\pi\i\left(\alpha_1+\alpha_2+\frac{bp}{2}\right)}
                \bigg(
                    \Theta_{\alpha_1+\alpha_2+bp,\,p}(\tau)\,\Bar{\Theta}_{\alpha_1-\alpha_2,\,p}(\Bar{\tau})
                    -
                    \Theta_{\alpha_1+\alpha_2+(b+1)p,\,p}(\tau)\,\Bar{\Theta}_{\alpha_1-\alpha_2+p,\,p}(\Bar{\tau})
                \bigg)\ .
    \end{aligned}
\end{align}
This expression correctly reproduces the result given in \cite{Kawabata:2023iss} when $C$ is self-dual.

\paragraph{Case with $p=N$}
In the case with $p=N$, the non-anomalous condition \eqref{non_anomalous} always holds while the other condition \eqref{all_one_condition} is satisfied if and only if the following holds:
\begin{align}\label{chi_condition(p=N)}
    (l_1\,{\bf 1}_{n},\, l_2\,{\bf 1}_{n})\notin C \qquad \text{and} \qquad (l_1\neq 0 \quad \text{or}\quad l_2\neq 0) \ .
\end{align}
In this case, we obtain
\begin{align}\label{lattice_theta(p=N)}
    \begin{aligned}
         \Theta_{\Gamma^a_{b}(C)}(\tau,\Bar{\tau})
            =
                \dfrac{1}{N}\sum_{\hat{a}\in\BZ_N}\omega_N^{-\hat{a}a}\,W_{C}\left(\{\Tilde{\psi}_{\alpha_1\alpha_2}(\hat a,b)\}\right)\ ,
    \end{aligned}
\end{align}
where 
\begin{align}
\label{psi_ab(p=N)}
    \begin{aligned}
        \Tilde{\psi}_{\alpha_1\alpha_2}(\hat a,b)
            =
                \omega_p^{\hat{a}(l_2\alpha_1+l_1\alpha_2+bl_1l_2)}\bigg( &\Theta_{\alpha_1+\alpha_2+(l_1+l_2)b,\,p}(\tau)\,\Bar{\Theta}_{\alpha_1-\alpha_2+(l_1-l_2)b,\,p}(\Bar{\tau})
                \\
            &\qquad
            +\Theta_{\alpha_1+\alpha_2+(l_1+l_2)b+p,\,p}(\tau)\,\Bar{\Theta}_{\alpha_1-\alpha_2+(l_1-l_2)b+p,\,p}(\Bar{\tau})\Bigg)\ .
    \end{aligned}
\end{align}

\paragraph{Remark:}\mbox{}\\
When $p=N$ and $(l_1\,{\bf 1}_{n},\, l_2\,{\bf 1}_{n})\in C$, the vector $\chi$ of the form \eqref{non_chiral_vector_chi} does not satisfy the condition \eqref{all_one_condition}.
However, there is another choice for the vector $\chi$ satisfying \eqref{all_one_condition} of the form:
\begin{align}
    \chi = \dfrac{1}{\sqrt{p}}\,(\,l_1\,{\bf 1}_{n},\, l_2\,{\bf 1}_{n}\,)\ , \qquad l_1,\,l_2\in\BZ_N\ .
\end{align} 
With this choice, the non-anomalous condition \eqref{non_anomalous} becomes
\begin{align}
\begin{aligned}
     n\,l_1\,l_2\in p^2\,\BZ\ .
\end{aligned}
\end{align}
The partition functions can be obtained by replacing $l_1 \to l_1/p,l_2 \to l_2/p$ in \eqref{psi_ab(p=N)}.

\subsubsection{Example: $c=1$ CFTs}
Let us consider the code CFT for $n=1$ generated by
\begin{align}
    G 
        =
            \begin{bmatrix}
               ~1~&~0~~
           \end{bmatrix} \,.
\end{align}
The $[2,1]_p$ code $C$ generated by $G$ is 
\begin{align}
\label{eq:codeword_c=1}
    C
        =
            \left\{(0,0),\, (1,0),\, \cdots,\, (p-1,0)\right\}\ .
\end{align}
Let us consider the case with $p=N$ and choose the vector $\chi = \sqrt{p}\,(0,l_2)$ for a non-zero integer $l_2\in \BZ_p$.

The complete weight enumerator of the code $C$ is
\begin{align}
    W_{C}
        =
            \sum_{i=0}^{p-1}\, x_{i0}\ .
\end{align}
Using \eqref{lattice_theta(p=N)} and \eqref{psi_ab(p=N)}, we obtain
\begin{align}\label{c=1_bosonic_PF}
    \begin{aligned}
         \Theta_{\Gamma^a_{b}}(\tau,\Bar{\tau})
        &=
            \Theta_{l_2^{p-2}a+l_2 b,\,p}(\tau)\,\Bar{\Theta}_{l_2^{p-2}a-l_2 b,\,p}(\Bar{\tau})
            +
            \Theta_{l_2^{p-2}a+l_2 b+p,\,p}(\tau)\,\Bar{\Theta}_{l_2^{p-2}a-l_2 b+p,\,p}(\Bar{\tau})\ ,
    \end{aligned}
\end{align}
where we used the Fermat's little theorem $l_2^{p-1}=1\ (\text{mod}\ p)$ for prime $p$ and $l_2\neq 0$.
It follows from \eqref{c=1_bosonic_PF} that the partition function of the code CFT is
\begin{align}
    \begin{aligned}
        Z_\CB(\tau,\Bar{\tau})
            &=
                \frac{1}{|\eta(\tau)|^2}\,\sum_{a\in \BZ_p}\, \Theta_{\Gamma^a_{0}}(\tau,\Bar{\tau}) \\
            &=
                \frac{2}{|\eta(\tau)|^2}\,\sum_{a\in \BZ_p}\,
                        \Theta_{a,\,p}(\tau)\,\Bar{\Theta}_{a,\,p}(\Bar{\tau}) \ ,
    \end{aligned}           
\end{align}
which shows that this is a compact boson theory of radius $R= \sqrt{2p}$ \cite{Ando:2024gcf}.

The orbifolded and bosonic theories are related by \eqref{orb_ferm_lattice_relations}, thus the theta functions for the orbifolded theory are given by
\begin{align}\label{c=1_orb_PF}
    \begin{aligned}
        \Theta_{\CO_\rho}[a, b]
            &=
                \Theta_{\Gamma^{\langle -\rho b\rangle}_{ \langle -a\rho^{p-2}\rangle}}(\tau,\Bar{\tau})\\
            &=
                \Theta_{l_2\rho^{p-2}a+l_2^{p-2}\rho b,\,p}(\tau)\,\Bar{\Theta}_{l_2\rho^{p-2}a-l_2^{p-2}\rho b,\,p}(\Bar{\tau})\\
            &\qquad \qquad
                +
                \Theta_{l_2\rho^{p-2}a+l_2^{p-2}\rho b+p,\,p}(\tau)\,\Bar{\Theta}_{l_2\rho^{p-2}a-l_2^{p-2}\rho b+p,\,p}(\Bar{\tau})\ .
    \end{aligned}
\end{align}
Comparing \eqref{c=1_bosonic_PF} with \eqref{c=1_orb_PF} and using the properties \eqref{theta_property}, we observe that the theta function of each sector is invariant under the orbifolding if $l_2$ and $\rho$ satisfy 
\begin{align}\label{c=1_self-duality_condition}
    \rho
        =
        l_2^2  \quad\pmod{p}\ .
\end{align}
Namely, we find a family of $c=1$ CFTs invariant under the $\BZ_p$-gauging that are labeled by prime $p$ and $l_2\in \BZ_p$.
This self-duality may be understood as follows; gauging the $\BZ_p$ shift symmetry in the $R=\sqrt{2p}$ compact boson yields the $R=\sqrt{2/p}$ theory, and performing the T-duality maps the latter to the original theory.
Note that this argument establishes self-duality only within the untwisted sector and does not yield any relations for the twisted sectors.
In contrast, our observation shows that the CFT becomes completely self-dual, i.e., both the untwisted and twisted sectors remain invariant, under the orbifolding when we gauge the theory with the parameter $\rho$ satisfying \eqref{c=1_self-duality_condition}.

Similarly, the parafermionized and bosonic theories are related by \eqref{orb_ferm_lattice_relations}, and the theta functions for the parafermionized theory are given by
\begin{align}
    \begin{aligned}
        \Theta_{\text{PF}_{\rho}}[a,b]
                &=
                    \Theta_{\Gamma^{\langle -a\rangle}_{ \langle a\rho^{p-2}-b\rangle}}(\tau,\Bar{\tau})\\
                &=
                    \Theta_{(l_2\rho^{p-2}-l_2^{p-2})a-l_2 b,\,p}(\tau)\,\Bar{\Theta}_{-(l_2\rho^{p-2}+l_2^{p-2})a+l_2 b,\,p}(\Bar{\tau})\\
                    &\qquad\qquad+
                    
                    \Theta_{(l_2\rho^{p-2}-l_2^{p-2})a-l_2 b+p,\,p}(\tau)\,\Bar{\Theta}_{-(l_2\rho^{p-2}+l_2^{p-2})a+l_2 b+p,\,p}(\Bar{\tau})\ .
    \end{aligned}
\end{align}

For any prime $p$, there always exists the solution to \eqref{c=1_self-duality_condition} given by $\rho=l_2=1$.
Let us examine this case for simplicity.
Then, the subsector of the Construction A lattice defined in \eqref{Lambda_a_b} becomes
\begin{align}
    \begin{aligned}
        \Lambda^a_{b}\left(C\right)
            =
                \left\{\dfrac{(a,b)}{\sqrt{p}}+\sqrt{p}\,(k_1,k_2)\,\bigg|\,k_1,k_2\in\BZ\right\}\ ,
    \end{aligned}
\end{align}
and the corresponding subsector of the momentum lattice is given by
\begin{align}\label{c_1_mom_lattice}
    \begin{aligned}
        \Gamma^a_{b}(C)
            =
                \left\{\dfrac{(a+b,a-b)}{\sqrt{2p}} + \sqrt{\dfrac{p}{2}}\,(k_1,k_2)\,\bigg|\,k_1,k_2\in\BZ,\,k_1=k_2~\text{mod}\ 2\right\}\ .
    \end{aligned}
\end{align}
By orbifolding, the subsector of the momentum lattice corresponding to the Hilbert space $(\CH^{\CO_1})_{b}^{a}$ is
\begin{align}\label{c_1_mom_lattice_orb}
    \begin{aligned}
        (\Gamma^{\CO_1})^a_{b}(C)&=\Gamma^{\langle-b\rangle}_{\langle-a\rangle}(C)\\
            &=
                \left\{\dfrac{(-a-b,a-b)}{\sqrt{2p}}+\sqrt{\dfrac{p}{2}}\,(k_1,k_2)\,\bigg|\,k_1,k_2\in\BZ,\,k_1=k_2~\text{mod}\ 2\right\}\ .
    \end{aligned}
\end{align}
Comparing \eqref{c_1_mom_lattice} and \eqref{c_1_mom_lattice_orb}, we find that the orbifolding does not change the spectrum of the theory as it acts as the reflection of the left-moving momentum $(p_L,p_R)\to (-p_L,p_R)$ on the momentum lattice.
Similarly, the subsector of the momentum lattice corresponding to the Hilbert space $(\CH^{\text{PF}_1})_{b}^{a}$ for the parafermionized theory is
\begin{align}\label{c_1_mom_lattice_paraferm}
    \begin{aligned}
         (\Gamma^{\text{PF}_1})^a_{b}(C)&=\Gamma^{\langle-a\rangle}_{\langle a-b\rangle}(C)\\
            &=
                \left\{\dfrac{(-b,-2a+b)}{\sqrt{2p}}+\sqrt{\dfrac{p}{2}}\,(k_1,k_2)\,\bigg|\,k_1,k_2\in\BZ,\,k_1=k_2~\text{mod}\ 2\right\}\ .
    \end{aligned}
\end{align}

To be more concrete, consider the case with $p=N=3$. 
In this case, the theta function for the subsector $\CH^{a}_{b}$ can be read off from \eqref{c=1_bosonic_PF}, and expanded in terms of $q$ and $\bar q$ as follows:
\begin{align}
\label{c=1_subsector_theta_function}
    \begin{aligned}
        \Theta_{\Gamma^{a}_{b}}
        =
         \begin{dcases}
             1+4\,q^{\frac{3}{4}}\Bar{q}^{\frac{3}{4}}+\cdots & (a,b)=(0,0)\\
             q^{\frac{1}{12}}\Bar{q}^{\frac{1}{12}}+q^{\frac{1}{3}}\Bar{q}^{\frac{1}{3}}(1+q+\Bar{q})+\cdots & (a,b)=(0,1),\, (0,2),\, (1,0),\, (2,0)\\
             2\,q^{\frac{1}{12}}\Bar{q}^{\frac{3}{4}}+q^{\frac{1}{3}}(1+q)+\cdots & (a,b)=(1,1),\, (2,2)\\
             2\,q^{\frac{3}{4}}\Bar{q}^{\frac{1}{12}}+\Bar{q}^{\frac{1}{3}}(1+\Bar{q})+\cdots & (a,b)=(1,2),\, (2,1)
         \end{dcases}
    \end{aligned}
\end{align}
Thus, the theta functions (which equal to the partition functions up to the $\eta$ function) in the untwisted sectors of the bosonic code CFT $(\CB)$, its orbifolded $(\CO_1)$ and parafermionized $(\text{PF}_1)$ theories with $\rho=1$ are given by
\begin{align}
    \begin{aligned}
        \Theta_{\CB}
            &:=
                \sum_{a\in\BZ_3}\Theta_{\Gamma^{a}_{0}}=
                    1 
                    + 
                        2\,q^{\frac{1}{12}}\Bar{q}^{\frac{1}{12}}
                    +
                        q^{\frac{1}{3}}\Bar{q}^{\frac{1}{3}}
                    +
                        2\,q^{\frac{3}{4}}\Bar{q}^{\frac{3}{4}}
                    +
                        q^{\frac{4}{3}}\Bar{q}^{\frac{1}{3}}
                    +
                        q^{\frac{1}{3}}\Bar{q}^{\frac{4}{3}}
                    +\cdots \ ,\\
        \Theta_{\CO_1}
           
           &:=
                 \sum_{a\in\BZ_3}\Theta_{\CO_{1}}[a, 0]
                 =
                    1 
                        + 
                            2\,q^{\frac{1}{12}}\Bar{q}^{\frac{1}{12}}
                        +
                            q^{\frac{1}{3}}\Bar{q}^{\frac{1}{3}}
                        +
                            2\,q^{\frac{3}{4}}\Bar{q}^{\frac{3}{4}}
                        +
                            q^{\frac{4}{3}}\Bar{q}^{\frac{1}{3}}
                        +
                            q^{\frac{1}{3}}\Bar{q}^{\frac{4}{3}}
                        +\cdots \ , \\
        \Theta_{\text{PF}_1}
            &:=
                \sum_{a\in\BZ_3}\Theta_{\text{PF}_1} [a,0]
                =
                    1
                        +
                            2\,\bar q^{\frac{1}{3}}
                        +
                            4\,q^{\frac{3}{4}}\Bar{q}^{\frac{1}{12}}
                        +
                            2\,\Bar{q}^{\frac{4}{3}}
                        +
                            4\,q^{\frac{3}{4}}\Bar{q}^{\frac{3}{4}}
                        + \cdots\ .
    \end{aligned}
\end{align}
We see that this CFT is self-dual under the orbifolding, i.e., $\Theta_\CB = \Theta_{\CO_1}$.
This is the expected result as the present $\BZ_{3}$-gauging satisfies the self-duality condition \eqref{c=1_self-duality_condition} with $\rho = l_2 = 1$.
Also, the $q$-expansion of $\Theta_{\text{PF}_1}$ is consistent with the spin selection rule \eqref{spin_selection_rule_orb_para}, which shows that the parafermionized theory contains the operators with fractional spin $s \in \BZ +\frac{2}{3}$ that can be identified with parafermions.

It is straightforward to repeat the same argument for $\rho=2$.
The orbifolded partition function is the same as before, $\Theta_{\CO_2} = \Theta_{\CO_1}$, as the untwisted sector of the orbifolded theory does not depend on the choice of $\rho$.
This means that the theory happens to be self-dual under the orbifolding with $\rho=2$ although there are no solutions to the condition \eqref{c=1_self-duality_condition} in this case.
On the other hand, the parafermionized partition function becomes
\begin{align}
\begin{aligned}
     \Theta_{\text{PF}_2}
        &:=
            \sum_{a\in\BZ_3}\Theta_{\text{PF}_2} [a,0]
        =
            1
                +
                    2\,q^{\frac{4}{3}}
                +
                    4\,q^{\frac{1}{12}}\Bar{q}^{\frac{3}{4}}
                +
                    2\,q^{\frac{1}{3}}
                +
                    4\,q^{\frac{3}{4}}\Bar{q}^{\frac{3}{4}}+\cdots\ ,
\end{aligned}
\end{align}
which contains the operators with fractional spin $s\in \BZ + \frac{1}{3}$ in accordance with the spin selection rule \eqref{spin_selection_rule_orb_para}, and is related to $\Theta_{\text{PF}_1}$ by the exchange $q\leftrightarrow\Bar{q}$.

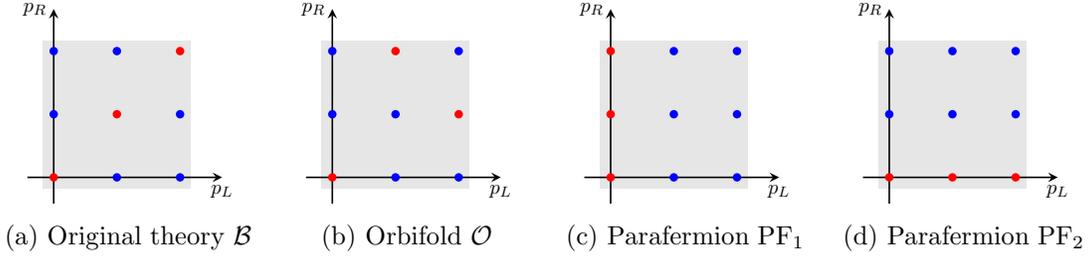
\begin{figure}
    \centering
    \begin{minipage}{0.23\textwidth}
    \begin{center}
        \begin{tikzpicture}[transform shape,scale=0.7]
    \fill[gray!20] (-0.2,-0.2)--(2.6,-0.2)--(2.6,2.6)--(-0.2,2.6)--cycle;

    \draw[semithick,->,>=stealth](-0.5,0)--(3.2,0) node [below] {$p_L$};
    \draw[semithick,->,>=stealth](0,-0.5)--(0,3.2) node[left] {$p_R$};

    \foreach \x in {0, 1.2, 2.4}
        \foreach \y in {0, 1.2, 2.4} {
            \ifdim \x pt = \y pt 
                \fill[red] (\x,\y) circle[radius=0.08cm];
            \else 
                \fill[blue] (\x,\y) circle[radius=0.08cm];
            \fi
        }
\end{tikzpicture}
    \end{center}
    \subcaption{Original theory $\CB$}
    \end{minipage}
    \begin{minipage}{0.23\textwidth}
    \begin{center}
       \begin{tikzpicture}[transform shape,scale=0.7]
    \fill[gray!20] (-0.2,-0.2)--(2.6,-0.2)--(2.6,2.6)--(-0.2,2.6)--cycle;

    \draw[semithick,->,>=stealth](-0.5,0)--(3.2,0) node [below] {$p_L$};
    \draw[semithick,->,>=stealth](0,-0.5)--(0,3.2) node[left] {$p_R$};

    \foreach \x in {0, 1.2, 2.4}
        \foreach \y in {0, 1.2, 2.4} {
            \pgfmathparse{int(mod(\x+\y,3))}            
            \ifnum \pgfmathresult=0 
                \fill[red] (\x,\y) circle[radius=0.08cm];
            \else 
                \fill[blue] (\x,\y) circle[radius=0.08cm];
            \fi
        }
\end{tikzpicture}
    \end{center}
    \subcaption{Orbifold $\CO$}
    \end{minipage}
    \begin{minipage}{0.23\textwidth}
    \begin{center}
        \begin{tikzpicture}[transform shape,scale=0.7]
    \fill[gray!20] (-0.2,-0.2)--(2.6,-0.2)--(2.6,2.6)--(-0.2,2.6)--cycle;

    \draw[semithick,->,>=stealth](-0.5,0)--(3.2,0) node [below] {$p_L$};
    \draw[semithick,->,>=stealth](0,-0.5)--(0,3.2) node[left] {$p_R$};

    \foreach \x in {0, 1.2, 2.4} {
        \foreach \y in {0, 1.2, 2.4} {
            \pgfmathparse{\x==0}
            \ifdim\pgfmathresult pt=1pt 
                \fill[red] (\x,\y) circle[radius=0.08cm];
            \else 
                \fill[blue] (\x,\y) circle[radius=0.08cm];
            \fi
        }
    }
\end{tikzpicture}
    \end{center}
    \subcaption{Parafermion $\mathrm{PF}_1$}
    \end{minipage}
    \begin{minipage}{0.23\textwidth}
    \begin{center}
        \begin{tikzpicture}[transform shape,scale=0.7]
    \fill[gray!20] (-0.2,-0.2)--(2.6,-0.2)--(2.6,2.6)--(-0.2,2.6)--cycle;

    \draw[semithick,->,>=stealth](-0.5,0)--(3.2,0) node [below] {$p_L$};
    \draw[semithick,->,>=stealth](0,-0.5)--(0,3.2) node[left] {$p_R$};

    \foreach \x in {0, 1.2, 2.4} {
        \foreach \y in {0, 1.2, 2.4} {
            \pgfmathparse{\y==0}
            \ifdim\pgfmathresult pt=1pt 
                \fill[red] (\x,\y) circle[radius=0.08cm];
            \else 
                \fill[blue] (\x,\y) circle[radius=0.08cm];
            \fi
        }
    }
\end{tikzpicture}
    \end{center}
    \subcaption{Parafermion $\mathrm{PF}_2$}
    \end{minipage}
    \caption{
    The fundamental regions of the momentum lattices of the untwisted sector in (a) the original bosonic theory $\CB$, (b) the $\BZ_{N=3}$-orbifolded theory $\CO$, (c) the parafermionized theory $\mathrm{PF}_1$ with $\rho=1$, and (d) the parafermionized theory $\mathrm{PF}_2$ with $\rho=2$.
    The original theory $\CB$ is based on the ternary $(p=3)$ code~\eqref{eq:codeword_c=1}. The other theories are gauged by the $\BZ_3$ symmetry specified by $\chi = \sqrt{3}\,(0,1)$.
    Within the fundamental region $\BF_3^2$ (shaded region), the red points represent the left- and right-moving momenta $(p_L,p_R)$ in the untwisted sectors $\CH_0$, $(\CH^{\CO})_0$, and $(\CH^{\mathrm{PF}_{1,2}})_0$. 
    Note that the untwisted sector of the orbifolded theory does not depend on the parameter $\rho$: $(\CH^{\CO_1})_0=(\CH^{\CO_2})_0$.
    }
    \label{fig:Z3gauging_code}
\end{figure}

Figure \ref{fig:Z3gauging_code} shows the spectra of the momentum lattices \eqref{c_1_mom_lattice}, \eqref{c_1_mom_lattice_orb} and \eqref{c_1_mom_lattice_paraferm} in the untwisted sectors of the code CFT and the orbifolded/parafermionized theories.
While the momentum lattices of the bosonic and its orbifolded theories look different, they are related by the reflection of the left momentum $p_L \to - p_L$ and give the same spectra.

\subsubsection{Example: $c=2$ CFTs}
As an example of a code CFT with $c=2$, we consider the $[4,2]_3$ code generated by
\begin{align}
    G 
        =
            \begin{bmatrix}
               ~1~ & ~0~ &  \,~~0~ &~1~~\\
               ~0~ & ~1~ &  \,-1~ &~0~~
           \end{bmatrix} .
\end{align}
In the case with $N=2$ and $(l_1,l_2)=(1,1)$, the fermionization from the $[4,2]_3$ code CFT gives rise to the $\CN=2$ supersymmetric CFT whose Ramond-Ramond partition function is constant \cite{Ando:2024gcf}. 

Here, we consider the $\BZ_3$-gauging with $N=3$ and $(l_1,l_2)=(1,1)$.
In this case, the theta functions \eqref{lattice_theta(p=N)} yield
\begin{align}
    \begin{aligned}
        \Theta_{\Gamma^a_{b}}
            =
             \begin{dcases}
                 1
                    +
                        2\,q^{\frac{1}{3}}\Bar{q}^{\frac{1}{3}}
                    +
                        4\,q^{\frac{1}{12}}\Bar{q}^{\frac{1}{12}}(q+\Bar{q})
                    +\cdots & (a,b)=(0,0)\\
                 2\,q^{\frac{1}{6}}\Bar{q}^{\frac{1}{6}}
                    +
                        q^{\frac{1}{3}}\Bar{q}^{\frac{1}{3}}
                    +
                        4\,q^{\frac{5}{12}}\Bar{q}^{\frac{5}{12}}
                    +
                        2\,q^{\frac{1}{12}}\Bar{q}^{\frac{1}{12}}(q+\Bar{q})+\cdots & (a,b)=(0,1),\,(0,2),\,(1,0),\,(2,0)\\
                 q^{\frac{1}{3}}
                 +
                    2\,q^{\frac{5}{12}}\Bar{q}^{\frac{1}{12}}
                +
                    2\,q^{\frac{1}{12}}\Bar{q}^{\frac{3}{4}}
                +
                    2\,q^{\frac{2}{3}}\Bar{q}^{\frac{1}{3}}
                +
                    4\,q^{\frac{1}{6}}\Bar{q}^{\frac{5}{6}}
                +\cdots & (a,b)=(1,1),\,(2,2)\\
                 \Bar{q}^{\frac{1}{3}}
                    +
                        2\,q^{\frac{1}{12}}\Bar{q}^{\frac{5}{12}}
                    +
                        2\,q^{\frac{3}{4}}\Bar{q}^{\frac{1}{12}}
                    +
                        2\,q^{\frac{1}{3}}\Bar{q}^{\frac{2}{3}}
                    +
                        4\,q^{\frac{5}{6}}\Bar{q}^{\frac{1}{6}}
                    +\cdots
                 & (a,b)=(1,2),\, (2,1) 
             \end{dcases}
    \end{aligned}
\end{align}
It follows that the partition functions of the untwisted sectors in the bosonic and orbifolded theory are given by
\begin{align}
    \begin{aligned}
        \Theta_{\CB}
            &=
        \Theta_{\CO}         
            =
                1
                    +
                        4\,q^{\frac{1}{6}}\Bar{q}^{\frac{1}{6}}
                    +
                        4\,q^{\frac{1}{3}}\Bar{q}^{\frac{1}{3}}
                    +
                        8\,q^{\frac{1}{6}}\Bar{q}^{\frac{1}{6}}
                    +
                        8\,q^{\frac{5}{12}}\Bar{q}^{\frac{5}{12}}(q+\Bar{q})\cdots \ ,
    \end{aligned}
\end{align}
where we drop the subscript $\rho$ for the orbifolded theory $\CO_\rho$ as the untwisted sector does not depend on the choice of $\rho$.
Thus, this theory is self-dual under the $\BZ_3$-gauging in the untwisted sector for both $\rho=1,2$.
On the other hand, the parafermionized theory depends on the choice of $\rho$ as follows:
\begin{align}
\begin{aligned}
     \Theta_{\text{PF}_1}
        &
            =
                1
                    +
                        2\,\Bar{q}^{\frac{1}{12}}(\Bar{q}^{\frac{1}{4}}
                    +
                        2\,q^{\frac{1}{12}}\Bar{q}^{\frac{1}{3}}
                    +
                        q^{\frac{1}{3}}\Bar{q}^{\frac{1}{4}}
                    +
                        2\,q^{\frac{4}{3}})
                    +\cdots \ ,\\
     \Theta_{\text{PF}_2}
        &
                =
                    1
                        +
                            2\,q^{\frac{1}{12}}(q^{\frac{1}{4}}
                        +
                            2\,q^{\frac{1}{3}}\Bar{q}^{\frac{1}{12}}
                        +
                            q^{\frac{1}{4}}\Bar{q}^{\frac{1}{3}}
                        +
                            2\,\Bar{q}^{\frac{4}{3}})
                        +\cdots\ .
\end{aligned}
\end{align}
We observe that the two parafermionic theories are related by the exchange $q \leftrightarrow \bar q$.

Next, we turn to the $p\neq N$ case. Let us consider the $[4,2]_2$ code generated by
\begin{align}
    G 
        =
            \begin{bmatrix}
               ~1~ & ~0~ &  ~0~ &~1~~\\
               ~0~ & ~1~ &  ~1~ &~0~~
           \end{bmatrix}\ .
\end{align}
From this generator matrix, we find that the code contains the codeword $({\bf 1}_2,{\bf 1}_2)$, allowing the choice of the $\BZ_3$-gauging specified by $\chi=\sqrt{2}\,({\bf 1}_2,{\bf 0}_2)$.\footnote{The other choice is $\sqrt{2}\,({\bf 0}_2,{\bf 1}_2)$, but the following results do not change.}
In this case, the partition functions in the untwisted sectors of the bosonic and orbifolded theory are given by
\begin{align}
    \begin{aligned}
        \Theta_{\CB}
            &=
                1
                    +
                        2\,q^{\frac{1}{9}}\Bar{q}^{\frac{1}{9}}
                    +
                        8\,q^{\frac{5}{18}}\Bar{q}^{\frac{5}{18}}
                    +
                        2\,q^{\frac{4}{9}}\Bar{q}^{\frac{4}{9}}
                    +
                        8\,q^{\frac{1}{2}}\Bar{q}^{\frac{1}{2}}
                    +
                        2\,(q+\Bar{q})+\cdots\ ,\\
        \Theta_{\CO}         
           &=
                1
                    +
                        8\,q^{\frac{1}{4}}\Bar{q}^{\frac{1}{4}}
                    +
                        16\,q^{\frac{1}{2}}\Bar{q}^{\frac{1}{2}}
                    +
                        4\,(q+\Bar{q})
                    +
                        \cdots\ .
    \end{aligned}
\end{align}
It turns out that this code CFT is not self-dual under the orbifolding.
On the other hand, the parafermionized theories are
\begin{align}
\begin{aligned}
     \Theta_{\text{PF}_1}
        &
            =
                1
                    +
                        4\,q^{\frac{1}{36}}\Bar{q}^{\frac{13}{36}}
                    +
                        2\,q^{\frac{1}{9}}\Bar{q}^{\frac{4}{9}}
                    +
                        4\,q^{\frac{25}{36}}\Bar{q}^{\frac{1}{36}}
                    +
                        2\,(q+q^{\frac{1}{2}}\Bar{q}^{\frac{1}{2}}+\Bar{q})
                    +\cdots \ ,\\
     \Theta_{\text{PF}_2}
        &
            =
                1
                    +
                        4\,q^{\frac{13}{36}}\Bar{q}^{\frac{1}{36}}
                    +
                        2\,q^{\frac{4}{9}}\Bar{q}^{\frac{1}{9}}
                    +
                        4\,q^{\frac{1}{36}}\Bar{q}^{\frac{25}{36}}
                    +
                        2\,(q+q^{\frac{1}{2}}\Bar{q}^{\frac{1}{2}}+\Bar{q})
                    +\cdots\ ,
\end{aligned}
\end{align}
from which we observe that the two theories are related by the exchange $q \leftrightarrow \bar q$.

\subsection{Chiral code CFTs}\label{ss:chiral-examples}

We consider chiral code CFTs constructed from binary doubly even self-dual codes.
Let $C$ be a doubly even self-dual code of length $n$.
When $C\subset\BF_2^n$ is self-dual, it always contains the all-ones vector $\mathbf{1}_n$.
The Construction A lattice $\Lambda(C) $ is even self-dual with respect to the standard Euclidean inner product $\cdot$.

Let us take a $\BZ_N$ symmetry specified by $\chi = \mathbf{1}_n/\sqrt{2}\in\Lambda(C)$.
One can check that this choice of $\chi$ satisfies the condition \eqref{characteristic_vector}.
On the other hand, the non-anomalous condition~\eqref{eq:non_anom} yields
\begin{equation}
    n\in 4\,N\,\BZ\ .
\end{equation}
Note that doubly even self-dual codes exist only when $n\in 8\,\BZ$.
Thus, the $\BZ_2$-gauging is possible for any $n\in 8\,\BZ$, while the $\BZ_N$-gaugings with odd prime $N$ are applicable only when $n\in 8\,N\,\BZ$.\par

In what follows, we will consider the case with $n=24$ for simplicity. 
In this case, it is known that there are nine doubly even self-dual codes $C_i~(i=1,2,\cdots, 9)$ 
\cite{pless1975classification-375,database}.
The classification of these codes and their Construction A lattices $\Lambda(C_i)$ are presented in table \ref{tab:code_classification}, where we follow the notation in \cite{conway2013sphere}.
\begin{table}[t]
    \centering
    \begin{tblr}{c|ccccccccc}
    \toprule
    $i$ & 1 & 2 & 3 & 4 & 5 & 6 & 7 & 8 & 9 \\ \midrule
    $C_i$ & $d_{12}^{~~2}$ & $d_{10}e_7^{~2}$ & $d_8^{~3}$ & $d_6^{~4}$ & $d_{24}$
    & $d_4^{~6}$ & $g_{24}$ & $d_{16}e_8$ & $e_8^{~3}$\\ [0.5em]
    $h_i$ & $\frac{5}{4}$ & $1$ & $\frac{3}{4}$ & $\frac{1}{2}$ & $\frac{11}{4}$ & $\frac{1}{4}$ & $0$ & $\frac{7}{4}$ & $\frac{7}{4}$\\ [0.5em]
    $\Lambda(C_i)$ & $D_{12}^{~~2}$ & $D_{10}E_7^{~2}$ & $D_8^{~3}$ & $D_6^{~4}$ & $D_{24}$
 & $D_4^{~6}$ & $A_1^{~24}$ & $D_{16}E_8$ & $E_8^{~3}$\\ [0.5em]
    $\Lambda^\CO_{\BZ_2}(C_i)$ & $D_{6}^{~4}$ & $D_{5}^{~2}A_7^{~2}$ & $D_4^{~6}$ & $A_3^{~8}$ & $D_{12}^{~~2}$
 & $A_1^{~24}$ & $\Lambda_{24}$ & $D_{8}^{~3}$ & $D_8^{~3}$\\ [0.5em]
 $\Lambda^\CO_{\BZ_3}(C_i)$ & $D_{6}^{~4}$/$D_6A_9^{~2}$ & $D_{5}^{~2}A_7^{~2}$ & $D_4^{~6}$/$D_4A_5^{~4}$ & $A_3^{~8}$ & $D_{12}^{~~2}$
 & $A_1^{~24}$ & $\Lambda_{24}$ & $D_{8}^{~3}$ & $D_8^{~3}$\\ \bottomrule
 \end{tblr}
    \caption{The classification of doubly even self-dual codes $C_i$ of length 24 and the correspondence between the codes and the associated lattices.
    The lattices $\Lambda^\CO_{\BZ_2}(C_i)$ of the orbifolded theories for the $\BZ_2$ symmetry are obtained in \cite{dolan1990conformal-df1,Dolan:1994st}.
    We identify the lattices $\Lambda^\CO_{\BZ_3}(C_i)$ for the $\BZ_3$ symmetry by calculating the lattice theta functions using the results in section \ref{ss:Z_N_gauged_code_CFTs}.
    }
    \label{tab:code_classification}
\end{table}
Each Construction A lattice falls into one of the twenty-four Niemeier lattices which exhaust even self-dual lattices of dimension 24.
With the above choice of the non-anomalous $\BZ_N$ symmetry, one can perform the orbifolding of the code CFTs associated with the nine codes $C_i$.
The orbifolded theories are also bosonic lattice CFTs whose momentum lattices $\Lambda^{\CO}(C_i)$ are one of the Niemeier lattices.

For the $\BZ_2$-orbifolding, the momentum lattices $\Lambda^{\CO}_{\BZ_2}(C_i)$ of the orbifolded theories are identified explicitly as in table \ref{tab:code_classification} in \cite{dolan1990conformal-df1,Dolan:1994st}, where twelve out of the twenty-four Niemeier lattices have been obtained as the momentum lattices through the Construction A and the orbifolding.

For the $\BZ_3$-orbifolding, we can calculate the lattice theta functions of the orbifolded theories by using the weight enumerator polynomials of the codes as in section \ref{ss:Z_N_gauged_code_CFTs}.
The weight enumerator polynomial of each code is given by \cite{nagaoka2023notetypeiicodes}
\begin{align}\label{WCi}
    W_{C_i}(\{x_{\alpha}\})=W_{C_{9}}(\{x_{\alpha}\})+6\,(4h_i-7)\, \Delta \ ,
\end{align}
where $\Delta:= x_0^4\,x_1^4(x_0^4-x_1^4)^4$, and $h_i$ is the number of codewords in $C_i$ of weight 4 divided by its length 24, whose value is listed in table \ref{tab:code_classification}. 
Since $C_9$ is the direct sum of three copies of the $[8,4]_2$ extended Hamming code $e_8$, its weight enumerator polynomial is given by $W_{C_{9}}(\{x_{\alpha}\})=(x_0^8+14\,x_0^4\,x_1^4+x_1^8)^3$.
Using \eqref{WCi}, the lattice theta functions of the orbifolded theories are obtained from the chiral version of the relation \eqref{lattice_theta_all_one}.

Comparing the lattice theta functions we obtain with those of the Niemeier lattices in \cite{nagaoka2017notes-e93}, we determine the lattices $\Lambda^\CO_{\BZ_3}(C_i)$ as in table \ref{tab:code_classification}.
Seven out of the nine lattices can be identified uniquely, while $\Lambda^\CO_{\BZ_3}(C_1)$ and $\Lambda^\CO_{\BZ_3}(C_3)$ each correspond to two possible Niemeier lattices as some Niemeier lattices have the same lattice theta functions \cite{nagaoka2017notes-e93}.

We can also construct the parafermionized theories of the code CFTs for the nine codes $C_i$ by the $\BZ_3$-gauging.
There are two types of parafermionization labeled by $\rho=1,2$.
The partition functions corresponding to each $\rho$ are summarized in table \ref{tab:chiral_parafermionization}.

\begin{table}[t]
    \centering
    \begin{tblr}{c|c|c}
        \toprule
         Code & $\Theta_{\text{PF}_1}$ & $\Theta_{\text{PF}_2}$\\ \midrule
        $d_{12}^{~~2}$ &  $1+2\,q^{\frac{2}{3}}+180\,q+17296\,q^{\frac{5}{3}}+61506\,q^2+\cdots$ & $1+180\,q+1566\,q^{\frac{4}{3}}+61506\,q^2+\cdots$\\ [0.1em]
        $d_{10}e_7^{~2}$ &  $1+2\,q^{\frac{2}{3}}+144\,q+17296\,q^{\frac{5}{3}}+62370\,q^2+\cdots$ & $1+144\,q+1566\,q^{\frac{4}{3}}+62370\,q^2+\cdots$\\ [0.1em]
        $d_8^{~3}$ &  $1+2\,q^{\frac{2}{3}}+108\,q+17296\,q^{\frac{5}{3}}+63234\,q^2+\cdots$ & $1+108\,q+1566\,q^{\frac{4}{3}}+63234\,q^2+\cdots$\\ [0.1em]
        $d_6^{~4}$ &  $1+2\,q^{\frac{2}{3}}+72\,q+17296\,q^{\frac{5}{3}}+64098\,q^2+\cdots$ &  $1+72\,q+1566\,q^{\frac{4}{3}}+64098\,q^2+\cdots$\\ [0.1em]
        $d_{24}$ &  $1+2\,q^{\frac{2}{3}}+396\,q+17296\,q^{\frac{5}{3}}+56322\,q^2+\cdots$ & $1+396\,q+1566\,q^{\frac{4}{3}}+56322\,q^2+\cdots$\\ [0.1em]
        $d_{4}^{~6}$ &  $1+2\,q^{\frac{2}{3}}+36\,q+17296\,q^{\frac{5}{3}}+64962\,q^2+\cdots$ & $1+36\,q+1566\,q^{\frac{4}{3}}+64962\,q^2+\cdots$\\ [0.1em]
        $g_{24}$ &  $1+2\,q^{\frac{2}{3}}+17296\,q^{\frac{5}{3}}+65826\,q^2+\cdots$ & $1+1566\,q^{\frac{4}{3}}+65826\,q^2+\cdots$\\ [0.1em]
        $d_{16}e_8$ &  $1+2\,q^{\frac{2}{3}}+252\,q+17296\,q^{\frac{5}{3}}+59778\,q^2+\cdots$ & $1+252\,q +1566\,q^{\frac{4}{3}}+59778\,q^2+\cdots$\\ [0.1em]
        $e_8^{~3}$ &  $1+2\,q^{\frac{2}{3}}+252\,q+17296\,q^{\frac{5}{3}}+59778\,q^2+\cdots$ & $1+252\,q+1566\,q^{\frac{4}{3}}+59778\,q^2+\cdots$\\ \bottomrule
        
    \end{tblr}
    \caption{The $\BZ_3$-parafermionized partition functions of the chiral code CFTs for the binary doubly even self-dual codes of length 24.}
    \label{tab:chiral_parafermionization}
\end{table}

\subsection{Code CFTs with $m>n$}\label{ss:asymmetric-examples}

We turn to code CFTs with asymmetric central charges $c = m$ and $\bar c = n$.
These CFTs can be constructed from binary $[m+n, \frac{m+n}{2}]_2$ codes generated by the matrices $G$ that satisfy the self-duality \eqref{code_self-duality} and doubly-evenness condition \eqref{code_doubly-even}.
Solving these conditions to obtain the general solutions is challenging, thus we will look for particular solutions of the form:
\begin{align}\label{CSS-like-code}
    G
        =
            \begin{bmatrix}
                \SG_\textsf{ch}    & 0        & 0 \\
                0                & \SH_\textsf{X}    & 0 \\
                0                & 0        & \SH_\textsf{Z}
            \end{bmatrix} \ ,
\end{align}
where $\SG_\textsf{ch}$ is an $(m-n)\times k$ binary matrix while $\SH_\textsf{X}$ and $\SH_\textsf{Z}$ are $n\times (n-k_\textsf{X})$ and $n\times (n-k_\textsf{Z})$ binary matrices, respectively.
The choice of this form is motivated by the CSS construction of quantum codes from a pair of classical codes \cite{Calderbank:1995dw,steane1996multiple}, so we will call codes generated by the matrix $G$ of the form the CSS-like codes.
Note that the three parameters $k,\,k_\textsf{X},\, k_\textsf{Z}$ are not independent, but subject to the relation:
\begin{align}\label{CSS_parameter_relation}
    k_\textsf{X} + k_\textsf{Z}
        =
            k
                +
                    \frac{3\,n - m}{2} \ .
\end{align}

Let us examine the self-duality condition \eqref{code_self-duality}.
Substituting the form \eqref{CSS-like-code} to \eqref{code_self-duality}, we obtain
\begin{align}\label{CSS_orthogonality_condition}
    \SG_\textsf{ch}\,\SG_\textsf{ch}^T = \SH_\textsf{X}\,\SH_\textsf{Z}^T  = 0 \qquad \pmod{2} \ .
\end{align}
Furthermore, the doubly-evenness condition \eqref{code_doubly-even} yields
\begin{align}\label{CSS-doubly-even-condition}
    \text{diag}\left( \SG_\textsf{ch}\,\SG_\textsf{ch}^T \right) = 0 \qquad \pmod{4} \ .
\end{align}
By regarding $\SG_\textsf{ch}$ as the generator matrix of an $[m-n,k]_2$ code $\SC_\textsf{ch}$, we find from \eqref{CSS_orthogonality_condition} and \eqref{CSS-doubly-even-condition} that $\SC_\textsf{ch}$ is a doubly even self-orthogonal code.
It follows from the self-orthogonality that the range of the parameter is restricted to $k \le \frac{m-n}{2}$.
Similarly, $\SH_\textsf{X}$ and $\SH_\textsf{Z}$ can be regarded as the check matrices of $[n, k_\textsf{X}]_2$ and $[n, k_\textsf{Z}]_2$ codes $\SC_\textsf{X}, \SC_\textsf{Z}$ that satisfy $\SC_\textsf{X}^\perp \subset \SC_\textsf{Z}$.\footnote{Here, the dual code $\SC_\textsf{X}^\perp$ is defined with respect to the Euclidean metric.}
The relation $\SC_\textsf{X}^\perp \subset \SC_\textsf{Z}$ implies that the two parameters $k_\textsf{X}$ and $k_\textsf{Z}$ are subject to the condition $n \le k_\textsf{X} + k_\textsf{Z}$.
Combining with \eqref{CSS_parameter_relation}, this condition yields $k\ge \frac{m-n}{2}$.
Since $k \le \frac{m-n}{2}$ due to the self-orthogonality of $\SC_\textsf{ch}$, we obtain $k= \frac{m-n}{2}$, which implies $\SC_\textsf{ch}$ is self-dual.
Hence, we find that the CSS-like code can be constructed from a doubly even self-dual $[m-n, \frac{m-n}{2}]_2$ code $\SC_\textsf{ch}$ and a pair of $[n, k_\textsf{X}]_2$ and $[n, n-k_\textsf{X}]_2$ codes $\SC_\textsf{X}, \SC_\textsf{Z}$ that satisfy $\SC_\textsf{X}^\perp \subset \SC_\textsf{Z}$.

Since the generator matrix \eqref{CSS-like-code} is block diagonal, the code CFT for the CSS-like code is a product of the chiral code CFT for the code $\SC_\textsf{ch}$ and the non-chiral code CFT for the code $\SC_\textsf{nc}$ generated by the matrix $
\big[
    \begin{smallmatrix}
        \SH_\textsf{X}      & 0 \\
        0                   &  \SH_\textsf{Z}
    \end{smallmatrix}
\big]
$.
Indeed, one can check by using \eqref{CWE} and \eqref{partition_enumerator_relation} that the partition function of the resulting CFT factorizes to the chiral and non-chiral parts:
\begin{align}\label{CSS-like_factorized_PF}
    Z
        =
            Z_\text{chiral}\cdot Z_\text{non-chiral} \ ,
\end{align}
where
\begin{align}
    \begin{aligned}
        Z_\text{chiral}
            &=
                \frac{1}{\eta(\tau)^{m-n}}\, W_{\SC_\textsf{ch}}\left( \{ \phi_\alpha\}\right) \ , \\
        Z_\text{non-chiral}
            &=
                \frac{1}{|\eta(\tau)|^{2n}}\, W_{\SC_\textsf{nc}}\left( \{ \psi_{ab}\}\right) \ .
    \end{aligned}
\end{align}
Note that the non-chiral part is modular invariant by itself while the chiral part is invariant under the modular $S$ transformation when $m-n\in 8\,\BZ_{>0}$ and invariant up to the phase $e^{\frac{2\pi\i\,(m-n)}{24}}$ under the modular $T$ transformation.

Next, we consider the gaugings of the code CFT by the $\BZ_N$ symmetry associated with the vector $\chi$ of the form \eqref{choice_of_chi}, which we reproduce here for later convenience:
\begin{align}\label{characteristic_vec_asymmetric}
    \chi
        =
            \sqrt{p}\,\left(\,l_0\,{\bf 1}_{m-n},\,l_1\,{\bf 1}_{n},\,l_2\,{\bf 1}_{n}\right) \ ,\qquad l_i\in\BZ_N ~~ (i=0,1,2)\ ,
\end{align}
For the $\BZ_N$ symmetry to act on the Hilbert space non-trivially, the vector $\chi$ must satisfy the condition \eqref{characteristic_vector}, but it is always met when $p$ and $N$ are coprime.
Since the CSS-like codes are binary codes, we will focus on the case with $p=2$ and odd prime $N$ below.
In this case, the non-anomalous condition \eqref{non_anom_asymmetric} for the $\BZ_N$ symmetry implies
\begin{align}\label{asymmetric_non-anomalous_condition}
    l_0^2\, (m-n) + 2\,l_1\,l_2\, n 
        \in
            N\,\BZ \ .
\end{align}
Since the present theory is a product of the chiral and non-chiral CFTs, the $\BZ_N$-gauging by the vector \eqref{characteristic_vec_asymmetric} acts as gauging the diagonal part of the $\BZ_N$ symmetries associated with the vectors $\chi_\textsf{ch} = \sqrt{p}\,l_0\,{\bf 1}_{m-n}$ and $\chi_\textsf{nc} = \sqrt{p}\,\left(l_1\,{\bf 1}_{n},\,l_2\,{\bf 1}_{n}\right)$ that act on the chiral and non-chiral CFTs, respectively.

We wish to gauge a $\BZ_N$ symmetry that acts non-anomalously on the whole CFT, but anomalously on the chiral and non-chiral parts.
To be concrete, we take $N=3$ and consider the two simple choices of parameters that solve the non-anomalous condition \eqref{asymmetric_non-anomalous_condition}:

\begin{table}[h]
    \centering
    \begin{tabular}{cccccc}
        \toprule
             & $m-n$ & $n$ & $l_0$ & $l_1$ & $l_2$ \\ \midrule
         (i) &  8    &  1  &  1     &   1   &   2  \\
        (ii) &  8    &  2  &  1     &   1   &   1  \\
        \bottomrule
    \end{tabular}
\end{table}

\noindent
In both cases, the chiral part is independent of the choice of the code $\SC_\textsf{ch}$ in the CSS-like code as the chiral CFT with central charge $8$ is unique \cite{Goddard:1989dp,dolan1990conformal-df1}.
This unique theory is known as the chiral $E_8$ CFT, whose partition function is given by (see e.g., \cite{Burbano:2021loy})
\begin{align}
    \begin{aligned}
        Z_{E_8}
            &=
                \frac{1}{2\,\eta(\tau)^8}\,\left[ \theta_2(\tau)^8 +  \theta_3(\tau)^8 +  \theta_4(\tau)^8\right] \\
            &=
                 \frac{1}{\eta(\tau)^8}\,\left[ 1 + 240\,q + 2160\,q^2 + 6720\,q^3 + \cdots\right] \ .
    \end{aligned}

\end{align}
On the other hand, the non-chiral part depends on the choice of a pair of the codes $\SC_\textsf{X}$ and $\SC_\textsf{Z}$.

For the case (i), the choice of the pair is unique up to the exchange of $\SC_\textsf{X}$ and $\SC_\textsf{Z}$ and given by $\SH_\textsf{X} = [\,1\,]$ and $\SH_\textsf{Z} = [\,0\,]$, where the latter should be interpreted as a $1\times 0$ matrix.
In this case, the bosonic partition function takes the factorized form of \eqref{CSS-like_factorized_PF}:
\begin{align}
    \Theta_{\CB}
            &=
                \big(1
                    +
                        240\,q
                    +
                        2160\,q^2
                    +                      
                    \cdots\big)
                \big(1
                    +
                        2\,q^{\frac{1}{8}}\Bar{q}^{\frac{1}{8}}
                    +
                        4\,q^{\frac{1}{2}}\Bar{q}^{\frac{1}{2}}
                    +
                        2\,q^{\frac{9}{8}}\Bar{q}^{\frac{1}{8}}
                    +
                        2\,q^{\frac{1}{8}}\Bar{q}^{\frac{9}{8}}
                    +
                        \cdots\big) \ .
\end{align}
On the other hand, the partition function of the orbifolded theory is given by
\begin{align}
    \begin{aligned}
        \Theta_{\CO}         
           &=
                1
                    +
                        2\,q^{\frac{2}{9}}\Bar{q}^{\frac{2}{9}}
                    +
                        2\,q^{\frac{25}{72}}\Bar{q}^{\frac{25}{72}}
                    +
                        84\,q
                    +
                        130\,q^{\frac{73}{72}}\Bar{q}^{\frac{1}{72}}
                    +
                        56\,q^{\frac{19}{18}}\Bar{q}^{\frac{1}{18}}
                    +    
                        \cdots\ .
    \end{aligned}
\end{align}
We find that the orbifolded theory does not factorize into the chiral $E_8$ CFT and a non-chiral CFT, hence it cannot be a code CFT built out of a CSS-like code.
For the parafermionized theories, we obtain the partition functions:
\begin{align}
\begin{aligned}
     \Theta_{\text{PF}_1}
        &
        
            =
                1
                    +
                        28\,q^{\frac{49}{72}}\Bar{q}^{\frac{1}{72}}
                    +
                        4\,q^{\frac{13}{18}}\Bar{q}^{\frac{1}{18}}
                    +
                        84\,q
                    +
                        2\,q^{\frac{25}{72}}\Bar{q}^{\frac{49}{72}}
                    +
                        130\,q^{\frac{8}{9}}\Bar{q}^{\frac{2}{9}}
                    
                    +\cdots \ ,\\
     \Theta_{\text{PF}_2}
        &
        
            =
                1
                    +
                        2\,q^{\frac{25}{72}}\Bar{q}^{\frac{1}{72}}
                    +
                        28\,q^{\frac{5}{9}}\Bar{q}^{\frac{2}{9}}
                    +
                        84\,q
                    +
                        28\,q^{\frac{49}{72}}\Bar{q}^{\frac{25}{72}}
                    +
                        2\,q^{\frac{2}{9}}\Bar{q}^{\frac{8}{9}}

                    +\cdots\ .
\end{aligned}
\end{align}
Both theories satisfy the spin selection rule \eqref{spin_selection_rule_orb_para}.

For the case (ii), there are a few choices of the pair:
\begin{enumerate}[label={(ii-\alph*)}]
    \item 
        \begin{align}
            \SH_\textsf{X} 
                =
                    \begin{bmatrix}
                        ~1~ & ~0~~\\
                        ~0~ & ~1~~
                    \end{bmatrix} \ , \qquad
            \SH_\textsf{Z} 
                =
                    \begin{bmatrix}
                        ~0~ & ~0~~
                    \end{bmatrix} \ .
        \end{align}
        For this choice, the partition functions in the untwisted sectors of the bosonic and orbifolded theory are given by
        \begin{align}
            \begin{aligned}
                 \Theta_{\CB}
                    &=
                        \big(1
                            +
                                240\,q
                            +
                                2160\,q^2
                            +                      
                            \cdots\big)
                        \big(1
                            +
                                4\,q^{\frac{1}{8}}\Bar{q}^{\frac{1}{8}}
                            +
                                4\,q^{\frac{1}{4}}\Bar{q}^{\frac{1}{4}}
                            +
                                8\,q^{\frac{1}{2}}\Bar{q}^{\frac{1}{2}}
                            +
                                \cdots\big)\ ,\\
                \Theta_{\CO}         
                   &=
                        1
                            +
                               8\,q^{\frac{1}{2}}\Bar{q}^{\frac{1}{2}}
                            +
                               968\,q^{\frac{5}{4}}\Bar{q}^{\frac{1}{4}}
                            +
                               1652\,q^{2}
                            +
                               1920\,q^{\frac{3}{2}}\Bar{q}^{\frac{1}{2}}
                            +
                               4\Bar{q}^{2}
                            +
                                \cdots\ ,
            \end{aligned}
        \end{align}
        and the parafermionized theories are
        \begin{align}
            \begin{aligned}
                 \Theta_{\text{PF}_1}
                    &=
                        1
                            +
                                316\,q^{\frac{9}{8}}\Bar{q}^{\frac{1}{8}}
                            +
                                8\,q^{\frac{5}{8}}\Bar{q}^{\frac{5}{8}}
                            +
                                324\,q^{\frac{5}{4}}\Bar{q}^{\frac{1}{4}}
                            +
                                352\,q^{\frac{5}{3}}\Bar{q}^{\frac{5}{3}}  
                            +
                                528\,q^{\frac{7}{6}}\Bar{q}^{\frac{1}{2}}
                            +
                                \cdots \ ,\\
                 \Theta_{\text{PF}_2}
                    &=
                        1
                            +
                                2\,q^{\frac{1}{4}}\Bar{q}^{\frac{1}{4}}
                            +
                                4\,q^{\frac{11}{24}}\Bar{q}^{\frac{1}{8}}
                            +
                                32\,q^{\frac{7}{12}}\Bar{q}^{\frac{1}{4}} 
                            +
                                84\,q
                            +
                                316\,q^{\frac{9}{8}}\Bar{q}^{\frac{1}{8}}
                            +    
                                8\,q^{\frac{5}{8}}\Bar{q}^{\frac{5}{8}}
                            +   
                                \cdots \ .
            \end{aligned}
        \end{align}

    \item 
        \begin{align}
            \SH_\textsf{X} 
                =
                    \begin{bmatrix}
                        ~1~ & ~0~~                    \end{bmatrix} \ , \qquad
            \SH_\textsf{Z} 
                =
                    \begin{bmatrix}
                        ~0~ & ~1~~
                    \end{bmatrix} \ .
        \end{align}

        With these choices, the partition functions of the bosonic code CFT, their orbifolded, and parafermionized theories are the same as in case (ii-a).

    \item 
        \begin{align}
            \SH_\textsf{X} 
                =
                    \begin{bmatrix}
                        ~1~ & ~1~~                    \end{bmatrix} \ , \qquad
            \SH_\textsf{Z} 
                =
                    \begin{bmatrix}
                        ~1~ & ~1~~
                    \end{bmatrix} \ .
        \end{align}

    With these choices, the partition functions of the bosonic and orbifolded theory are given by
        \begin{align}
            \begin{aligned}
                \Theta_{\CB}
                    &=
                \Theta_{\CO}         
                  =
                        \big(1
                            +
                                240\,q
                            +
                                2160\,q^2
                            +                      
                            \cdots\big)
                        \big(1
                            +
                                8\,q^{\frac{1}{4}}\Bar{q}^{\frac{1}{4}}
                            +
                                16\,q^{\frac{1}{2}}\Bar{q}^{\frac{1}{2}}
                            +    
                                4\,q
                            +
                                4\,\Bar{q}
                            +
                                \cdots\big)\ .
            \end{aligned}
        \end{align}
        Hence, this theory is self-dual under the $\BZ_3$-gauging in the untwisted sectors. 
        On the other hand, the partition functions of the parafermionized theories are
        \begin{align}
            \begin{aligned}
                 \Theta_{\text{PF}_1}
                    &=
                        1
                            +
                                4\,q^{\frac{1}{4}}\Bar{q}^{\frac{1}{4}}
                            +
                                30\,q^{\frac{2}{3}}
                            +
                                86\,q
                            +
                                4\,\Bar{q}
                            +
                                384\,q^{\frac{11}{12}}\Bar{q}^{\frac{1}{4}}  
                            +
                                648\,q^{\frac{5}{4}}\Bar{q}^{\frac{1}{4}}
                            +
                                8\,q^{\frac{1}{4}}\Bar{q}^{\frac{5}{4}}
                            +
                                \cdots \ ,\\
                 \Theta_{\text{PF}_2}
                    &=
                        1
                            +
                                2\,q^{\frac{1}{3}}
                            +
                                4\,q^{\frac{1}{4}}\Bar{q}^{\frac{1}{4}}
                            +
                                64\,q^{\frac{7}{12}}\Bar{q}^{\frac{1}{4}}
                            +
                                86\,q
                            +
                                4\,\Bar{q}
                            +
                                430\,q^{\frac{4}{3}}
                            +
                                224\,q^{\frac{5}{6}}\Bar{q}^{\frac{1}{2}}
                            +
                                8\,q^{\frac{1}{3}}\Bar{q}
                            +
                                
                                \cdots \ .
            \end{aligned}
        \end{align}
        
\end{enumerate}

\section{CFTs self-dual under $\BZ_N$-orbifolding}\label{ss:self-dual}

In recent studies, theories self-dual under orbifolding have played a distinguished role as they are shown to have duality defects that generate non-invertible symmetries by the half gauging construction \cite{Choi:2021kmx,Shao:2023gho}.
These works motivate us to identify a class of codes that give rise to CFTs self-dual under $\BZ_N$-orbifolding through our construction.
In this section, we will apply the $\BZ_N$-gaugings of code CFTs developed in the previous sections to search for such theories.

Let us focus on non-chiral code CFTs constructed from the CSS code over $\BF_p$ for prime $p$:
\begin{align}
    C_{\text{CSS}}=\left\{(c_1,c_2)\in\BF_{p}^{n}\times\BF_{p}^{n}\,\Big|\,c_1\in \SC^{\perp},\,c_2\in \SC\right\} \ ,
\end{align}
where $\SC$ is a code of length $n$ over $\BF_{p}$ and $\SC^\perp$ is the dual code of $\SC$ with respect to the Euclidean inner product.
The corresponding Construction A lattice takes the form:
\begin{align}
\label{construction_A_CSS}
\begin{aligned}
     \Lambda(C_{\text{CSS}})&=\left\{\, \left(\frac{c_1 + p\,k_1}{\sqrt{p}},\frac{c_2 + p\,k_2}{\sqrt{p}}\right)\, \bigg|\, c_1\in  \SC^{\perp},\, c_2\in \SC\ , ~ k_1,k_2 \in \BZ^{n} \right\}\\
    &=\Lambda(\SC^{\perp})\times\Lambda(\SC)\ ,
\end{aligned}
\end{align}
where $\Lambda(\SC^{\perp})$ and $\Lambda(\SC)$ are the Construction A lattices of the codes $\SC^{\perp}$ and $\SC$,  respectively.\par

In what follows, we set $p=N$ and consider the orbifolded theory by the $\BZ_p$ symmetry characterized by a vector $\chi=\sqrt{p}\,({\bf 0}_n,{\bf 1}_n)$.
As we saw in section \ref{ss:non-chiral-examples}, the vector $\chi$ must satisfy \eqref{chi_condition(p=N)} so that it acts as a $\BZ_p$ symmetry non-trivially.
In the present setup, it yields the following condition on the code $\SC$:
\begin{align}
\label{CSS_allone_condition}
    {\bf 1}_n\notin \SC\ .
\end{align}

Next, we turn to the lattice corresponding to the untwisted sector of the orbifolded theory given by
\begin{align}
\label{orbifold_ConA_lattice}
     \Lambda^{\CO}
        :=
            \bigoplus_{a\in\BZ_p}\left(\Lambda^{\CO_{\rho}}\right)^{a}_{0}
        =
            \bigoplus_{a\in\BZ_p}\Lambda^{0}_{a}(C_{\text{CSS}}) \ ,
\end{align}
where we used the relation \eqref{correspondence_hilbert_spaces} between the bosonic and orbifolded theories in the second equality.\footnote{We omit the subscript $\rho$ for the lattice of the orbifolded theory $\CO_\rho$ as the untwisted sector does not depend on the choice.}
We now identify the elements in the sector $\Lambda^{0}_{a}(C_{\text{CSS}})$.
It follows from \eqref{Lambda_a_b_shift_relation} that the sector $\Lambda^{0}_{a}(C_{\text{CSS}})$ is obtained by shifting $\Lambda^0_0(C_\text{CSS})$ as
\begin{align}
    \Lambda^{0}_{a}(C_{\text{CSS}})
        =
            \Lambda^{0}_{0}(C_{\text{CSS}})+\dfrac{a}{N}\chi\ ,
\end{align}
where we used $\ell=0$ that follows from \eqref{def_ell} and $\chi^2=0$ for the present choice of the $\BZ_p$ symmetry.
The subsector $\Lambda^{0}_{0}(C_{\text{CSS}})$ is the charge $0$ sector of the Construction A lattice \eqref{ConstructionA}, thus its vector $\mu=(\mu_1,\mu_2)\in\Lambda^{0}_{0}(C_{\text{CSS}})$ is expressed by
\begin{align}
    \begin{aligned}
        \mu_1=\dfrac{c_1 + p\,k_1}{\sqrt{p}},\qquad\mu_2=\frac{c_2 +  p\,k_2}{\sqrt{p}} \ , \qquad (k_1,k_2\in\BZ^n) \ ,
    \end{aligned}
\end{align}
where $(c_1,c_2)\in C_{\text{CSS}}$ satisfies
\begin{align}
    \begin{aligned}
        \mu\circledcirc_{\Lambda}\chi = 0 \pmod{p}\ ,
    \end{aligned}
\end{align}
which follows from the condition \eqref{grading_condition}.
On the other hand, the direct calculation shows
\begin{align}
    \begin{aligned}
        \mu\circledcirc_{\Lambda}\chi={\bf 1}_n\cdot c_1 \pmod{p}\ .
    \end{aligned}
\end{align}
Thus, the elements of $\Lambda^{0}_{0}(C_{\text{CSS}})$ are the Construction A lattice vectors built from the codewords $(c_1, c_2)\in C_\text{CSS}$ subject to the condition ${\bf 1}_n\cdot c_1 =0 \pmod{p}$. 
To recapitulate the above discussion, the lattice corresponding to the untwisted sector of the orbifolded theory $\Lambda^{\CO}$
is given by
\begin{align}
\label{orbifold_CSS}
\begin{aligned}
     \Lambda^{\CO}
        &=
            \bigoplus_{a\in\BZ_p}\Bigg\{\, \left(\frac{c_1 + p\,k_1}{\sqrt{p}},\frac{c_2 + a\,{\bf 1}_n+ p\,k_2}{\sqrt{p}}\right)\, \bigg|\, c_1\in \SC^{\perp},~ c_2\in \SC\ , ~ \\
        &\hspace{3cm} k_1,k_2 \in \BZ^{n} \ , ~ {\bf 1}_n\cdot c_1=0 \pmod{p}\Bigg\} \ .
\end{aligned}
\end{align}

Now, we move on to searching for code CFTs invariant under the orbifolding by the $\BZ_p$ symmetry up to T-duality.
Reminding that T-duality acts as $p_R \to -p_R$ on the momentum lattice, it acts as swapping $\SC$ and $\SC^\perp$ on the Construction A lattice as seen from \eqref{lambda-transformation}.
Hence, if the lattice $\Lambda^\CO$ takes the same form as the original lattice \eqref{construction_A_CSS} up to T-duality, namely,
\begin{align}
\label{T-dual_CSS}
    \Lambda^{\CO}=\Lambda(\SC)\times\Lambda(\SC^{\perp})\ ,
\end{align}
then the code CFTs become self-dual under the orbifolding.
Comparing \eqref{orbifold_CSS} and \eqref{T-dual_CSS}, we find the two conditions on the code $\SC$:
\begin{align}
\label{right_condition}
    \SC^{\perp}= \SC\oplus\langle\bf{1}_n\rangle \ ,
\end{align}
and 
\begin{align}
\label{left_condition}
    \SC=\left\{c'\in \SC^{\perp}\,\big|\,{\bf 1}_n\cdot c'=0\pmod{p}\right\}\ .
\end{align}
These conditions are not independent, but \eqref{left_condition} is indeed the necessary condition for \eqref{right_condition} as seen below. 
Let $c'\in \SC^{\perp}$ be a codeword satisfying \eqref{right_condition}. Then, there exists $c\in \SC$ and $\lambda\in\BZ_p$ such that 
\begin{align}
\label{CSS_dual_code}
    c' = c+\lambda\,{\bf 1}_n\ .
\end{align}
Then, the inner product between $c'\in \SC^{\perp}$ written in the form of \eqref{CSS_dual_code} and ${\bf 1}_n$ is
\begin{align}\label{one_dot_cprime}
\begin{aligned}
    {\bf 1}_n \cdot c' &= {\bf 1}_n \cdot  c + \lambda~{\bf 1}_n\cdot{\bf 1}_n\\
    &=\lambda~n\quad \pmod{p}\ ,
\end{aligned}
\end{align}
where we used ${\bf 1}_n\cdot c = 0~(\text{mod} ~p)$ for $c\in \SC$ in the second equality as seen from \eqref{right_condition}.
If $n \in p\,\BZ$, then for all $c' \in \SC^{\perp}$, ${\bf 1}_n \cdot c' = 0\pmod{p}$. This implies ${\bf 1}_n\in(\SC^{\perp})^{\perp}= \SC$, which contradicts the condition \eqref{CSS_allone_condition}, so we must have $n\notin p\,\BZ$. 
In this case, it follows from \eqref{one_dot_cprime} that the condition ${\bf 1}_n \cdot c' = 0~(\text{mod}~p)$ is equivalent to $\lambda=0~(\text{mod}~p)$.
Thus, if the condition ${\bf 1}_n \cdot c' = 0~(\text{mod}~p)$ holds for $c'\in \SC^\perp$, then $c'\in \SC$.
Conversely, we see from \eqref{CSS_dual_code} and \eqref{one_dot_cprime} that $c'\in \SC$ implies ${\bf 1}_n \cdot c' = 0~(\text{mod}~p)$.
This proves our statement that if the code $\SC$ satisfies \eqref{right_condition}, then \eqref{left_condition} is automatically satisfied.\footnote{
For an $[n,k]$ code $\SC$, \eqref{right_condition} implies $n=2k+1$, and hence $n$ must be an odd number that is not multiple of $p$.
(The latter condition follows from the discussion around \eqref{one_dot_cprime}.
An alternative proof goes as follows.
Suppose ${\bf 1}_n \in \SC$, then from the condition ${\bf 1}_n\cdot c = 0~\pmod{p}$, one finds ${\bf 1}_n\cdot {\bf 1}_n = n = 0~\pmod{p}$ and $n$ has to be a multiple of $p$.
By taking the contraposition, we prove the statement that ${\bf 1}_n \not\in \SC$ when $n$ is not a multiple of $p$.
)
}
Therefore, we can construct code CFTs that are self-dual under the $\BZ_p$ gauging from $[n,\frac{n-1}{2}]_p$ codes satisfying \eqref{right_condition}.\par
We can find a family of the $[n,\frac{n-1}{2}]_p$ codes that satisfy \eqref{right_condition} within the class of cyclic codes. 
An $[n,k]_p$ cyclic code $\SC$ is defined by a generator polynomial $g(x)=\sum_{i=0}^{n-k}g_ix^i\in\BF_{p}[x]$ that divides the polynomial $x^n-1$. The generator matrix $G_\SC$ of the code $\SC$ is given by
\begin{align}
    \begin{aligned}
        G_\SC=\begin{bmatrix}
            ~g_0~ & ~g_1~ & ~g_2~ & ~\cdots~  &  ~g_{n-k}~ & ~0~ & ~\cdots~ & ~0~~\\
            ~0~   & ~g_0~ & ~g_1~ & ~\cdots~  & ~g_{n-k-1}~ & ~g_{n-k}~& ~0~ &~\cdots~\\
                  & \vdots& \vdots& ~\vdots~  & ~\vdots~ & ~\vdots~ &~~&\\
                  0 & ~\cdots~ & g_0 & & ~\cdots~ & && g_{n-k}
        \end{bmatrix}\ .
    \end{aligned}
\end{align}
The generator polynomial $g^{\perp}(x)\in\BF_{p}[x]$ of the dual code $\SC^{\perp}$ can be obtained from $g(x)$ as (see e.g.,\cite{huffman2003fundamentals}[Theorem 4.2.7])
\begin{align}
    g^{\perp}(x)=\dfrac{x^k h(x^{-1})}{h(0)}\in\BF_{p}[x]\ ,
\end{align}
where $h(x)=(x^n-1)/g(x)$ and $k$ is degree of $g(x)$, and it is known that $\SC$ is self-orthogonal if and only if $g^{\perp}(x)\mid g(x)$. 
In order to find the cyclic code that satisfies the condition \eqref{right_condition}, we must choose the generator polynomial $g(x)$ satisfying the following conditions:
\begin{itemize}
    \item $g(1)=0$ and $g^\perp(1)\neq 0$  \qquad(equivalent to ${\bf 1}_n\in \SC^{\perp}$ and ${\bf 1}_n\not\in \SC$)
    \item The degree of $g(x)$ is $\frac{n+1}{2}$ \qquad(implies $k=\frac{n-1}{2}$)
    \item $g^{\perp}(x)\mid g(x)$ \qquad(implies $\SC\subset \SC^{\perp}$)
\end{itemize}
We can directly construct the generator polynomial $g(x)$ when $p= 1\pmod{n}$. Let  $a$ be one of the primitive elements of $\BF_p$ (i.e., $\BF_p=\{0,1=a^0,a^1,\ldots,a^{p-2}\}$). Then, $a^{i\,\frac{p-1}{n}}\ (i=0,1,\cdots,n-1)$ belong to $\BF_{p}$ and represent all solutions of $x^{n}-1=0$ in $\BF_{p}$. Therefore, $x^n-1$ can be fully factorized over $\BF_{p}$ as follows:
\begin{align}
    \begin{aligned}
        x^n-1&=\prod_{i=0}^{n-1}\left(x-a^{i\,\frac{p-1}{n}}\right)\\
        &=(x-1)\prod_{i=1}^{\frac{n-1}{2}}\left(x-a^{i\,\frac{p-1}{n}}\right)\left(x-a^{-i\,\frac{p-1}{n}}\right)\ ,
    \end{aligned}
\end{align}
where $a^{-i\,\frac{p-1}{n}}\coloneqq a^{(n-i)\,\frac{p-1}{n}}\ (i\geq 1)$ is multiplicative inverse of $a^{i\frac{p-1}{n}}$. Based on this factorization, we define the generator polynomial $g(x)$ which satisfies $g(x)\mid (x^n-1)$ by
\begin{align}
\label{self_dual_CFT_generator_polynomial}
    g(x)=(x-1)\prod_{i=1}^{\frac{n-1}{2}}\left(x-a^{i\,\frac{p-1}{n}}\right)\ .
\end{align}
Then, we can check that $g(x)=(x-1)g^{\perp}(x)$ and this $g(x)$ satisfies the three conditions above to define the cyclic code satisfying \eqref{right_condition}. 
From the above discussion, if $p= 1\pmod{n}$, self-dual code CFTs can be constructed from the cyclic codes. 
Since $n$ is an odd number, there exist $p$ and $n$ such that $p= 1\pmod{n}$ if and only if $p$ cannot be written in the form $p=2^m+1$ for $m\in\BN$ (Fermat prime).

\paragraph{Example:}
Let us consider a simple example with $ n = 3 $ and $p = 7$. The polynomial $x^3 - 1$ can be factorized in $\mathbb{F}_7$ as
\begin{align}
    x^3 - 1 = (x - 1)(x - 2)(x - 4)\ .
\end{align}
We have two choices of the generator polynomials $g_1(x)$ and $g_2(x)$ of the form \eqref{self_dual_CFT_generator_polynomial}:
\begin{align}
    \begin{aligned}
        g_1(x)&=(x-1)(x-2)&\qquad g_2(x)&=(x-1)(x-4)\\
        &=x^2+4x+2& &=x^2+2x+4
    \end{aligned}
\end{align}
The difference arises from the choice of the primitive elements in $\BF_7$.
These polynomials $g_1(x)$ and $g_2(x)$ define the two cyclic $[3,1]_7$ codes $\SC_1$ and $\SC_2$ with generator matrices:
\begin{align}
\label{Example_CSS_generator_matrix}
    \begin{aligned}
        G_{\SC_1}=\begin{bmatrix}
        ~1~ & ~2~ & ~4~~
    \end{bmatrix}\ ,\qquad G_{\SC_2}=\begin{bmatrix}
        ~1~ & ~4~ & ~2~~
    \end{bmatrix}\ .
    \end{aligned}
\end{align}
These matrices satisfy the self-orthogonality conditions, $G_{\SC_1} G_{\SC_1}^{T}=G_{\SC_2} G_{\SC_2}^{T}=0\pmod{7}$, and we can verify that these codes satisfy the condition \eqref{right_condition} that gives rise to the self-dual code CFT.
From \eqref{Example_CSS_generator_matrix}, one can observe that the only difference between $\SC_1$ and $\SC_2$ is a permutation of the coordinates. Consequently, the partition functions corresponding to the code CFTs of $\SC_1$ and $\SC_2$ have no difference.
The partition functions in the untwisted sectors of the bosonic code CFT and its orbifolded theory are given by
\begin{align}
    Z_{\CB}=Z_{\CO}=\dfrac{1}{|\eta{(\tau)}|^6}\left[1+2q^{\frac{3}{28}}\Bar{q}^{\frac{3}{28}}+6q^{\frac{5}{28}}\Bar{q}^{\frac{5}{28}}+6q^{\frac{3}{14}}\Bar{q}^{\frac{3}{14}}+\cdots\right]\ .
\end{align}
In the above discussion, we have shown that the code CFTs become self-dual when $x^n-1$ can be factorized into $n$ monic linear polynomials over $\BF_p$.
However, in other cases, there exist cyclic codes whose code CFTs are self-dual.
For example, $x^7-1$ is factorized over $\BF_{11}$ as
\begin{align}
    x^7-1=(x-1)(x^3+5x^2+4x+10)(x^3+7x^2+6x+10)\ ,
\end{align}
and let us set $g(x)=(x-1)(x^3+5x^2+4x+10)$. Then, we can check that $g(x)=(x-1)\,g^{\perp}(x)$ and this $g(x)$ gives a $[7,3]_{11}$ cyclic code satisfying \eqref{right_condition}.

\section{Discussion}\label{ss:discussion}

In this paper, we considered gaugings of bosonic lattice CFTs by non-anomalous $\BZ_N$ symmetries.
In section \ref{ss:ZN_bosonic}, we derived the spin selection rule for operators in a $\BZ_N$-charge-twisted sector for general bosonic CFTs and read off the correspondence between the Hilbert spaces of the bosonic and orbifolded/parafermionized theories from the transformation laws of the torus partition functions under $\BZ_N$-gaugings.
In section \ref{ss:ZN-gauging-lattice-CFTs}, we formulated $\BZ_N$-gaugings in bosonic lattice CFTs as modifications of the momentum lattices by a lattice vector that specifies a $\BZ_N$ symmetry.
We expressed the condition for a $\BZ_N$ symmetry to be non-anomalous in terms of the associated vector and showed that the torus partition functions of the orbifolded and parafermionized theories in a charge-twist sector can be written by lattice theta functions.
We then applied our formulation to code CFTs in section \ref{ss:codeCFT} and derived the $\BZ_N$-gauged partition functions expressed by the weight enumerator polynomials of the underlying codes.
Our formulation was illustrated by concrete examples in section \ref{ss:Examples} and also leveraged to identify a class of CSS codes that give rise to a family of self-dual bosonic CFTs under $\BZ_N$-orbifolding in section \ref{ss:self-dual}.

We have focused on bosonic lattice CFTs based on even self-dual lattices obtained from codes through Construction A in this paper.
Alternatively, it is also possible to leverage another class of codes to yield odd self-dual lattices, correspondingly fermionic lattice CFTs as in~\cite{Gaiotto:2018ypj,Kawabata:2023nlt} for chiral cases. 
For theories with central charges $c=m$ and $\bar c = n$, one can generalize the discussion in section \ref{ss:code_to_lattice} to fermionic cases as follows:
\begin{itemize}
    \item For an odd prime $p$, the Construction A lattice $\Lambda(C)$ is odd self-dual with respect to $\circledcirc_\Lambda$ if and only if the code $C\subset\BF_p^{m+n}$ is self-dual with respect to $\circledcirc_\Lambda$.
    \item For  $p=2$, the Construction A lattice $\Lambda(C)$ is an odd self-dual lattice with respect to $\circledcirc_\Lambda$ if and only if the code $C\subset\BF_2^{m+n}$ is singly even self-dual with respect to $\circledcirc_\Lambda$, where singly even codes are defined as
    \begin{align}
        ^{\exists}c= (c_0,c_1,c_2)\in C\ ,\quad c\circledcirc_\Lambda c =  c_0^2 + 2\,c_1\cdot c_2\in 4\,\BZ+2 \ .
    \end{align}
\end{itemize}
As well as the bosonic CFTs we have discussed, the fermionic lattice CFTs can possess a non-anomalous $\BZ_N$ symmetry in addition to the fermion parity symmetry. 
Our gauging technique employing lattice modifications may be extended to perform a $\BZ_N$-gauging of fermionic lattice CFTs, potentially leading to the orbifolding and parafermionization of fermionic CFTs (see \cite{Gaiotto:2018ypj} for the study of the orbifoldings of fermionic chiral CFTs).

The parafermionic CFTs constructed in the present work arise from bosonic lattice CFTs through specific lattice modifications. 
Meanwhile, fermionic lattice CFTs can be obtained from bosonic lattice CFTs by modifying the momentum lattice as discussed in \cite{Kawabata:2023usr,Kawabata:2023iss,Ando:2024gcf}, and also directly from odd self-dual lattices (see e.g., \cite{Gaiotto:2018ypj,Kawabata:2023nlt} for chiral cases and \cite{Kawabata:2025hfd} for non-chiral cases).
This motivates us to speculate that a certain class of lattices could directly yield parafermionic CFTs. 
While parafermionic minimal models have been classified in \cite{Yao:2020dqx}, extending this classification to a broader range of theories remains an open direction for future exploration.
Identifying such a class of lattices would be valuable for the classification of parafermionic CFTs, at least for small central charges.

The $\BZ_N$-orbifolding and parafermionization in two dimensions have been explored from the viewpoint of symmetry topological field theories (SymTFTs) in \cite{Duan:2023ykn}, where the gauging procedures are described in terms of topological boundary conditions in three-dimensional BF theories with level $N$.
Meanwhile, the recent work \cite{Kawabata:2025hfd} has constructed fermionic code CFTs from bosonic abelian Chern-Simons (CS) theories by imposing topological boundary conditions determined by odd self-dual codes in  (see also \cite{Kawabata:2023iss,Barbar:2023ncl,Dymarsky:2024frx} for the SymTFT construction of bosonic code CFTs).
The SymTFT perspective on the $\BZ_N$-gaugings in \cite{Duan:2023ykn} suggests a possible extension of the framework in \cite{Kawabata:2025hfd} to the construction of parafermionic code CFTs from bosonic Abelian CS theories via suitable topological boundary conditions.

\acknowledgments
We are grateful to T.\,Okuda and S.\,Yahagi for valuable discussions and collaborations in related works.
The work of K.\,A. was supported by JST SPRING, Grant Number JPMJSP2138.
The work of K.\,K. was supported by FoPM, WINGS Program, the University of Tokyo, and by JSPS KAKENHI Grant-in-Aid for JSPS fellows Grant No.\,23KJ0436.
The work of T.\,N. was supported in part by the JSPS Grant-in-Aid for Scientific Research (B) No.\,24K00629, Grant-in-Aid for Scientific Research (A) No.\,21H04469, and
Grant-in-Aid for Transformative Research Areas (A) ``Extreme Universe''
No.\,21H05182 and No.\,21H05190.

\bibliographystyle{JHEP}
\bibliography{QEC_CFT_prime_power}
\end{document}